\documentclass[12pt,draftclsnofoot,onecolumn]{IEEEtran}

\usepackage[utf8]{inputenc} 
\usepackage[T1]{fontenc}
\usepackage{url}              
\usepackage{cite}             

\usepackage[cmex10]{amsmath}  
\usepackage{mleftright}       
\mleftright                   

\usepackage{graphicx}         
\usepackage{booktabs}         

\newtheorem{lem}{Lemma}
\newtheorem{coro}{Corollary}

\newtheorem{thm}{Theorem}
\newtheorem{define}{Definition}
\newtheorem{remark}{Remark}

\usepackage{amsfonts}

\begin{document}

\title{\LARGE Compute-Forward Multiple Access for Gaussian MIMO Channels}


%

\author{Lanwei Zhang,~\IEEEmembership{Student Member,~IEEE,}
        Jamie Evans,~\IEEEmembership{Senior Member,~IEEE,}
        and~Jingge Zhu,~\IEEEmembership{Member,~IEEE.}
}

\maketitle

\begin{abstract}
Compute-forward multiple access (CFMA) is a multiple access transmission scheme based on Compute-and-Forward (CF) which allows the receiver to first decode linear combinations of the transmitted signals and then solve for individual messages. This paper extends the CFMA scheme to a two-user Gaussian multiple-input multiple-output (MIMO) multiple access channel (MAC). 
We propose the CFMA serial coding scheme (SCS) and the CFMA parallel coding scheme (PCS) with nested lattice codes. We first derive the expression of the achievable rate pair for MIMO MAC with CFMA-SCS. We prove a general condition under which CFMA-SCS can achieve the sum capacity of the channel. Furthermore, this result is specialized to single-input multiple-output (SIMO) and $2$-by-$2$ diagonal MIMO multiple access channels, for which more explicit sum capacity-achieving conditions on power and channel matrices are derived. We construct an equivalent SIMO model for CFMA-PCS and also derive the achievable rates. Its sum capacity achieving conditions are then analysed. Numerical results are provided for the performance of CFMA-SCS and CFMA-PCS in different channel conditions. In general, CFMA-PCS has better sum capacity achievability with higher coding complexity.
\end{abstract}


\section{Introduction}\label{sec_intro}
Due to the broadcast nature of the medium, interference occurs when multiple transmitters send their signals simultaneously in a wireless network, where it can be problematic for a receiver to recover the desired message. The traditional orthogonal transmission schemes avoid interference by allocating the transmission resources orthogonally to every transmitter. For example, the time-division multiple access (TDMA) scheme divides each transmitted signal into different time slots. In this case, each time slot only contains the signal from a single transmitter thus no interference occurs. However, these methods suffer from a diminishing rate, especially when many transmitters send messages simultaneously \cite{Tse05fundamentalsof}. Treating interference as noise (TIN) is another scheme practically used to handle interference. However, when the interference level is high, the transmission rate will be low due to the small effective signal-to-noise ratio (SNR).

Unlike orthogonal schemes or TIN, physical-layer network coding (PLNC)\cite{Zhang06PLNC} exploits interference directly by recovering a function of the received signals. Compute-and-forward (CF)\cite{Nazer11CF} is a linear PLNC scheme, which allows the receiver to decode an integer linear combination of the messages from multiple transmitters. Nested lattice codes are used in CF because any integer linear combination of lattice points is still a lattice point thus single-user decoders are enough for the decoding process. The CF scheme has been applied to various wireless communication channels\cite{Nazer11CF,Zhu17,nazer16expand,Hong13,Ordentlich17,Zhan14,He18,Lyu19} and has been shown to achieve the best performance among all the transmission schemes at that time. 

The multiple access channel (MAC) is one of the most common wireless channel models, which is sometimes called up-link channel in the cellular network. When applying CF to MAC, we require the same amount of linear combinations as the transmitters to recover the individual message of each user. Compute-forward multiple access (CFMA) \cite{Zhu17} is a multiple access scheme that allows unequal rates for each user and achieves the MAC capacity region when SNR is large enough. However, most of the current lattice codes-based multiple access schemes consider the single-input single-output (SISO) case, such as \cite{Zhu17}, or the single-input multiple-output (SIMO) case, such as \cite{Zhan14,He18,nazer16expand}. In this paper, we consider a multi-user multiple-input multiple-output (MIMO) MAC, where each user is equipped with multiple transmit antennas. We apply the CFMA framework to this model and propose the corresponding transmission schemes.  

\subsection{Related work}

As shown in \cite{Erez05}, lattices are good for noise tolerance and quantization. Nested lattice codes are applied to different AWGN point-to-point channels \cite{Gamal04,Erez04,Hindy15,Campello16}. In \cite{Erez04}, it is shown that with lattice encoding and decoding, the capacity is achieved for the AWGN channel (with single antenna). The authors in \cite{Hindy15,Campello16} apply lattice-based codes in fading channels to achieve the ergodic capacity. The class of LAttice Space–Time (LAST) codes has been proposed in \cite{Gamal04} and achieves the optimal diversity–multiplexing tradeoff of MIMO Channels. In \cite{Ordentlich15}, the authors show that with precoded nested lattice codes, they can achieve the optimal MIMO capacity to within a constant gap. 

In the CF scheme \cite{Nazer11CF}, the receiver decodes integer linear combinations of the received lattice codewords. It can either forward the linear combination to the next receiver, thus working as a relay, or collect enough linear combinations to recover each message. Based on the latter idea, the integer-forcing receiver was proposed in \cite{Zhan14} as a generalization of the classical zero-forcing receiver. Given enough linearly independent equations, individual messages can be easily resolved. It shows a great improvement in rate due to the flexible choice of the equation coefficients. Successive computing for the linear combinations is then applied in \cite{Zhu17,nazer16expand,He18}, where the former decoded linear combinations will be ``cancelled" in the decoding process for the latter linear combinations. In this way, the computation rate is further improved. The original CF scheme is extended in \cite{Zhu17,nazer16expand} by allowing unequal message rates for different users. Based on this extension, it is shown that for a two-user SISO multiple access channel (MAC), CFMA \cite{Zhu17} can achieve the whole capacity pentagon without time sharing if the SNR is large enough. We note that in \cite{nazer16expand}, the authors proposed an expanded CF scheme for SIMO MAC and show that their scheme can achieve the sum capacity given some power constraints and channel conditions. However, their results do not take an explicit form. As one of the contributions in this paper, we give more concise sum capacity-achieving conditions for the two-user SIMO channels.

The channel models studied with the CF scheme in the literature usually contain SISO MAC \cite{Zhan14,Zhu17} and SIMO MAC \cite{nazer16expand,He18}. It has been shown that the CF scheme with lattice codes is capacity achievable with some channel conditions in both cases. However, very few results are directly related to the MIMO MAC case. We are interested in how the CF scheme with lattice codes can be applied to the MIMO MAC case and whether it is still capacity achievable.  

\subsection{Paper contributions}
This paper extends the CFMA scheme to a two-user Gaussian MIMO MAC, where the transmitters and the receiver have multiple antennas. We propose two coding schemes based on CFMA. In the CFMA Serial Coding Scheme (SCS), each transmitter is equipped with a single lattice codebook, and a codeword is divided and transmitted in part by each transmit antenna. We also propose the CFMA Parallel Coding Scheme (PCS) where each transmit antenna is equipped with an individual codebook. The capacity achievability is discussed for both schemes. The paper contributions are as follows.

\begin{itemize}
    \item As a generalization of the result in \cite{Zhu17}, we analyze CFMA-SCS and give the analytical expressions of the achievable rate pair.
    \item We show that the MIMO MAC sum capacity is achievable with CFMA-SCS given certain polynomial-form conditions. Although the conditions are not always satisfiable for all power constraints and channel conditions, we provide some capacity-achieving guarantees when the power constraint is large enough for some special cases, including single-input multiple-output (SIMO) MAC and diagonal MIMO MAC. 
    \item We analyze the CFMA Parallel Coding Scheme (PCS) and its achievable rate. General channel conditions are found to achieve the sum capacity. 
    \item Numerical examples are provided to show the capacity achievability for CFMA-SCS and CFMA-PCS, where they are compared in different channel conditions. We observe that in general CFMA-PCS has better sum capacity achievability than CFMA-SCS. However, CFMA-SCS has lower coding complexity. In addition, CFMA-SCS seems to be more sensitive to the line-of-sight (LoS) channel strength.
\end{itemize}

\section{System Model}\label{sec_model}
We consider a two-user Gaussian MIMO MAC where each user is equipped with $t$ transmit antennas and the receiver is equipped with $r$ receive antennas. The channel coefficients are assumed fixed and known to transmitters and receivers during the transmissions. They are denoted as $\textbf{H}_l$ with dimension $r\times t$ for $l=1,2$. We use $\textbf{x}_{l,s}\in \mathbb R^t$ to denote the channel input of user $l$ at time $s$. At time $s$, the channel output is given by
\begin{equation}
    \label{eq_channel_output_MIMO_single_channel_use}
    \textbf{y}_s = \sum_{l=1}^2 \textbf{H}_l \textbf{x}_{l,s} + \textbf{z}_s,
\end{equation}
where $\textbf{z}_s$ is the additive white Gaussian noise with zero mean and unit variance. Let $n$ denote the total number of channel uses. The channel input for each user is constrained by an input covariance matrix $\textbf{K}_l = \mathbb{E}[\textbf{x}_{l,s}\textbf{x}_{l,s}^T] \succeq \textbf{0}$ with dimension $t\times t$, which satisfies the power constraint $tr(\textbf{K}_l) \leq P$ for $l=1,2$, where $tr(\cdot)$ refers to the trace of a matrix. 

We will extend the CFMA scheme in \cite{Zhu17} from SISO MAC to the two-user MIMO MAC. In the SISO case, each user is naturally equipped with a single lattice codebook. In the MIMO case, since each user has multiple antennas, we could either construct a single codebook for each user, or a single codebook for each transmitting antenna. Based on this observation, we propose two different coding schemes in this paper. For brevity, we will denote the first scheme, where each user has a single codebook, as the CFMA serial coding scheme (SCS) and the second scheme, where each transmit antenna has an individual codebook, as the CFMA parallel coding scheme (PCS). In the two-user case, with CFMA-SCS we will have two codebooks whose codeword length is $tn$ while CFMA-PCS requires $2t$ codebooks in total with codeword length $n$.

\section{The CFMA serial coding scheme (SCS)}\label{sec_SCS}

In this section, we will study the serial coding scheme for a two-user MIMO MAC. Recall that we will construct a $tn$-length codebook for each user. To simplify the presentation, we apply a concatenating channel model \cite{Gamal04}\cite{Lin11}, where all the $n$ symbols are concatenated into a $tn$-dimensional vector, i.e., $\textbf{x}_l = (\textbf{x}_{l,1}^T,\ldots,\textbf{x}_{l,n}^T)^T$. The concatenating channel output for all the channel uses is given by
\begin{equation}
    \label{eq_channel_output_MIMO_vector}
    \textbf{y} = \sum_{l=1}^2 \bar{\textbf{H}}_l \textbf{x}_l + \textbf{z},
\end{equation}
where $\textbf{y}, \textbf{z}\in\mathbb{R}^{rn}$ and $\bar{\textbf{H}}_l := \textbf{I}_n \otimes \textbf{H}_l$ are $rn \times tn$ block diagonal matrices. Note that $\textbf{I}_n$ is the $n$-dimensional identity matrix, and $\otimes$ denotes the Kronecker product. The channel input covariance matrix of each user in this concatenating model is given by $\bar{\textbf{K}}_l := \textbf{I}_n \otimes \textbf{K}_l$, for $l = 1,2$.

\subsection{Transmission scheme for CFMA-SCS}\label{sec_coding_SCS}
The codebook of user $l$ for $l=1,2$ is constructed by a $tn$-dimensional nested lattice and is denoted by
\begin{equation}\label{eq_codebook_MIMO}
    \mathcal{C}_l = \Lambda_l^F\cap \mathcal{V}_l^C,
\end{equation}
where the fine lattice $\Lambda_l^F$ is chosen to be good for AWGN channel coding in the sense of \cite{Erez05}, and $\mathcal{V}_l^C$ is the Voronoi region of the coarse lattice $\Lambda_l^C$, which is chosen to be good for quantization in the sense of \cite{Erez05}. We denote the second moment of the coarse lattice as
\begin{equation}\label{eq_second_moment_MIMO}
    \sigma^2(\Lambda_l^C) = \frac{1}{tn\text{Vol}(\mathcal{V}_l^C)}\int_{\mathcal{V}_l^C} ||\textbf{x}||^2d\textbf{x} = \beta_l^2,\quad\beta_l\in\mathbb{R}.
\end{equation}
With this codebook, the message rate of user $l$ is given by
\begin{equation}\label{eq_message_rate_MIMO}
    r_l = \frac{\log|\mathcal{C}_l|}{n} = \frac{1}{n}\log\frac{\text{Vol}(\mathcal{V}_l^C)}{\text{Vol}(\mathcal{V}_l^F)}.
\end{equation}

Given a codeword $\textbf{t}_l$, we can define $\textbf{c}_l = [\textbf{t}_l/\beta_l+\textbf{d}_l]\mod \Lambda_l^C /\beta_l$, where $\textbf{d}_l$ is a dither vector which is uniformly distributed in the region $\Lambda_l^C /\beta_l$, and $\mod$ is the modulo operation. We can also write $\textbf{c}_l$ as
\begin{equation}
    \label{eq_c_l}
    \textbf{c}_l = (\textbf{t}_l/\beta_l+\textbf{d}_l) - \mathcal{Q}_{\Lambda_l^C /\beta_l}(\textbf{t}_l/\beta_l+\textbf{d}_l),
\end{equation}
where $\mathcal{Q}_\Lambda$ is the quantization operation over the lattice $\Lambda$. Readers are referred to \cite{Nazer11CF,Zhu17} for more details about the nested lattice codes. The channel input is generated as
\begin{equation}
    \label{eq_channel_input_MIMO}
    \textbf{x}_l = \bar{\textbf{B}}_l \textbf{c}_l,
\end{equation}
where $\bar{\textbf{B}}_l: = \textbf{I}_n \otimes \textbf{B}_l$ is a $tn\times tn$ matrix such that $\textbf{K}_l = \textbf{B}_l \textbf{B}_l^T$. Since $\textbf{K}_l$ is always positive semi-definite, there always exists such a $\textbf{B}_l$, for example, by Cholesky decomposition. However notice that since we do not require $\textbf{B}_l$ to be triangular in our construction, this composition is not unique.

To recover the individual messages, the receiver will decode two linearly independent integer combinations of the lattice codewords. For each linear combination, the receiver will apply a proper equalization matrix to the channel output. The equalized channel output can be then written as the summation of the desired linear combination and the effective noise. Since any integer linear combination of the lattice points is still a lattice point, we can apply a single-user decoder to decode the desired linear combination. In this paper, we will apply the ambiguity lattice decoder of \cite{Loeliger97}, which is defined in Definition~\ref{def_ambuiguity_lattice_decoder}. 
\begin{define}[Ambiguity lattice decoder]\label{def_ambuiguity_lattice_decoder}
    For a $n$-dimensional lattice $\Lambda\subset \mathbb{R}^n$, the ambiguity lattice decoder with a decision region $\mathcal{E}\in\mathbb{R}^n$ gives the result $\textbf{u}\in\mathbb{R}^n$ if the obtained signal  $\textbf{r}\in\Lambda$ can be written in the form $\textbf{r} = \textbf{u} + \textbf{z}$ with a unique lattice codeword $\textbf{u}\in\Lambda$ and the noise $\textbf{z}\in \mathcal{E}$. Otherwise, it produces an ambiguity error.
\end{define} 

After decoding the first linear combination (successfully), the receiver can ``cancel'' it in the decoding process for the second linear combination, i.e., successive decoding. In the next section, we will discuss how the two linear combinations are computed and derive the expressions of achievable rate pairs.

\subsection{Achievable Rates with CFMA-SCS}\label{sec_rate_SCS}
In this section, we will give the achievable rate pair for recovering the individual messages by decoding two linearly independent integer linear combinations with coefficients $\textbf{a}\in\mathbb{Z}^2$ and $\textbf{b}\in\mathbb{Z}^2$, respectively. Note that $|\textbf{M}|$ stands for the determinant of the square matrix $\textbf{M}$ and $Pr(E)$ is the probability of event $E$.
\begin{thm}\label{thm_rate_pair}
For a two-user MIMO Gaussian MAC given the channel matrices $\textbf{H}_1,\textbf{H}_2$ and the precoding matrices $\textbf{B}_1,\textbf{B}_2$, with CFMA-SCS, the following rate pair is achievable
\begin{equation}\label{eq_achievable_rate_MIMO_MAC_CFMA}
    R_l = \left\{\begin{array}{ll}
        r_l(\textbf{a},\boldsymbol{\beta}), & b_l = 0 \\
        r_l(\textbf{b}|\textbf{a},\boldsymbol{\beta}), & a_l = 0 \\
        \min\{r_l(\textbf{a},\boldsymbol{\beta}),\;r_l(\textbf{b}|\textbf{a},\boldsymbol{\beta})\}, & \text{otherwise}
    \end{array}\right.
\end{equation}
for any linearly independent $\textbf{a},\textbf{b}\in\mathbb{Z}^2$ and $\boldsymbol{\beta}\in\mathbb{R}^2$ if $r_l(\textbf{a},\boldsymbol{\beta})\geq 0$ and $r_l(\textbf{b}|\textbf{a},\boldsymbol{\beta}) \geq 0$ for $l = 1,2$, where
\begin{align}
    r_l(\textbf{a},\boldsymbol{\beta}) = & \frac{1}{2} \log 
    \frac{\beta_l^{2t} \left|\textbf{I}_{r} + \textbf{H}_1 \textbf{K}_1\textbf{H}_1^T + \textbf{H}_2 \textbf{K}_2\textbf{H}_2^T\right|}
    {\left| \textbf{M} \right|}, \label{eq_achievable_rate_MIMO_first} \\
    r_l(\textbf{b}|\textbf{a},\boldsymbol{\beta}) = & \frac{1}{2} \log
    \frac{\beta_l^{2t} {\left| \textbf{M} \right|}}
    {(\tilde a_1 \tilde b_2 - \tilde a_2 \tilde b_1)^{2t}}, \label{eq_achievable_rate_MIMO_second} \\
    \textbf{M} = & (\tilde a_1^2 + \tilde a_2^2)\textbf{I}_{t} + (\tilde a_1 \textbf{B}_2^T \textbf{H}_2^T - \tilde a_2 \textbf{B}_1^T \textbf{H}_1^T)(\tilde a_1 \textbf{H}_2 \textbf{B}_2 - \tilde a_2 \textbf{H}_1 \textbf{B}_1) \label{eq_Mh}
\end{align}
with $\tilde a_l = a_l \beta_l, \tilde b_l = b_l \beta_l$ for $l=1,2$.
\end{thm}

\begin{IEEEproof}
The coefficients of the first linear combination are denoted as $\textbf{a} = (a_1,a_2)^T$. Let the receiver compute
\begin{align*}
    \textbf{y}_1^\prime = & \textbf{W} \textbf{y} -\sum_l a_l\beta_l \textbf{d}_l\\
    = & \textbf{W}(\sum_l\bar{\textbf{H}}_l \bar{\textbf{B}}_l \textbf{c}_l + \textbf{z})-\sum_l a_l\beta_l \textbf{d}_l -\sum_l a_l\beta_l \textbf{c}_l+\sum_l a_l\beta_l \textbf{c}_l\\
    = & \underbrace{\textbf{W}\textbf{z} + \sum_l (\tilde a_l\textbf{I}_{tn}-\textbf{W}\bar{\textbf{H}}_l\bar{\textbf{B}}_l )(-\textbf{c}_l)}_{\bar{\textbf{z}}_1}-\sum_l a_l\beta_l \textbf{d}_l\\
    & + \sum_l a_l\beta_l(\textbf{t}_l/\beta_l+\textbf{d}_l) - a_l\beta_l\mathcal{Q}_{\Lambda_l^C /\beta_l}(\textbf{t}_l/\beta_l+\textbf{d}_l) \\
    = & \bar{\textbf{z}}_1 + \sum_l a_l \underbrace{\left(\textbf{t}_l - \mathcal{Q}_{\Lambda_l^C }(\textbf{t}_l+\beta_l\textbf{d}_l)\right)}_{\tilde{\textbf{t}}_l},
\end{align*}
where $\textbf{W}$ is a $tn\times rn$ equalization matrix to be determined later. The last equality holds because $\mathcal{Q}_\Lambda(\beta \textbf{x}) = \beta \mathcal{Q}_{\Lambda/\beta}(\textbf{x})$ for any lattice $\Lambda$ and any nonzero real $\beta$. It is enough to recover the message from $\tilde{\textbf{t}}_l$ since $\tilde{\textbf{t}}_l$ and the codeword $\textbf{t}_l$ belong to the same coset of $\Lambda_f^C$ \cite{Zhu17,Gamal04}. 

We will next show with the ambiguity lattice decoding defined in Definition~\ref{def_ambuiguity_lattice_decoder}, the achievable rate to decode $\sum_l a_l \tilde{\textbf{t}}_l$ is given by
\begin{equation}\label{eq_achievable_rate_general}
    r_{l}^{(1)} = \max_{\textbf{W}}\quad \frac{1}{2n} \log^+ \frac{|\beta_l^2\textbf{I}_{tn}|}{|\boldsymbol{\Sigma}_1(\textbf{W})|},
\end{equation}
where $|\cdot|$ refers to the determinant, and $\boldsymbol{\Sigma}_1(\textbf{W}) = \textbf{W}\textbf{W}^T + \sum_l (\tilde a_l\textbf{I}_{tn}-\textbf{W}\bar{\textbf{H}}_l\bar{\textbf{B}}_l)(\tilde a_l\textbf{I}_{tn}-\textbf{W}\bar{\textbf{H}}_l\bar{\textbf{B}}_l)^T$. For brevity, we may use $\boldsymbol{\Sigma}_1$ to refer to $\boldsymbol{\Sigma}_1(\textbf{W})$. Recall that we are considering an ensemble of $tn$-dimensional nested lattices $\{\Lambda_l^C \subseteq \Lambda_l^F\}$ for $l=1,2$ with message rate given by (\ref{eq_message_rate_MIMO}). We let $\Lambda_C = \Lambda_l^C / \beta_l$ thus $\sigma^2(\Lambda_C) = 1$ from (\ref{eq_second_moment_MIMO}). We further let $\Lambda_F = \Lambda_l^F$ for $l=1,2$ with Voronoi region $\mathcal{V}_F$. To prove the achievable rate in (\ref{eq_achievable_rate_general}), we will apply the ambiguity lattice decoder in Definition~\ref{def_ambuiguity_lattice_decoder} to the lattice $\Lambda_F$ with the decision region
\begin{equation}
    \label{eq_decision_region}
    \mathcal{E}_{tn,\eta} = \{\textbf{z}\in\mathbb{R}^{tn}: |\textbf{Q}\textbf{z}|^2\leq tn(1+\eta)\},
\end{equation}
for some $\eta>0$ where $\textbf{Q}\in\mathbb{R}^{tn\times tn}$ satisfies $\textbf{Q}^T\textbf{Q} = \boldsymbol{\Sigma}_1^{-1}$. The decomposition of $\boldsymbol{\Sigma}_1^{-1}$ is always
possible since $\boldsymbol{\Sigma}_1$ is positive-definite. From Definition~\ref{def_ambuiguity_lattice_decoder}, the decoding error happens in two cases. The first case is when the effective noise leaves the decision region centered at the correct codeword. The second case is when the effective noise stays in the decision region, but an ambiguity error occurs. The ambiguity error is the event $\mathcal{A}$ that the received point belongs to $(\mathcal{E}_{tn,\eta} + \textbf{u}_1) \cap (\mathcal{E}_{tn,\eta} + \textbf{u}_2)$ for some pair of distinct lattice points $\textbf{u}_1,\textbf{u}_2\in \Lambda_F$. For the equivalent channel $\textbf{y}_1^\prime = \bar{\textbf{z}}_1 + \textbf{u}$ where $\textbf{u} =  \sum_l a_l \tilde{\textbf{t}}_l \in \Lambda_F$, with the union bound, the decoding error probability is upper-bounded by
\begin{equation}
    \label{eq_error_prob}
    P_e \leq (\bar{\textbf{z}}_1 \notin \mathcal{E}_{tn,\eta}) + Pr(\mathcal{A}|\bar{\textbf{z}}_1 \in \mathcal{E}_{tn,\eta}),
\end{equation}
By taking the expectation over the ensemble of random
lattices, from \cite[Theorem 4]{Loeliger97} we can upper bound the second term of the RHS of (\ref{eq_error_prob}) thus the average error probability is upper bounded by 
\begin{equation}
    \label{eq_average_error_prob}
    \bar P_e \leq Pr(\bar{\textbf{z}}_1 \notin \mathcal{E}_{tn,\eta}) + (1+\delta)\frac{\text{Vol}(\mathcal{E}_{n,\eta})}{\text{Vol}(\mathcal{V}_F)},
\end{equation}
for any $\delta>0$. The term $\text{Vol}(\mathcal{E}_{tn,\eta})$ refers to the volume of the decision region which can be represented by
\begin{equation}
    \label{eq_vol_decision_region}
    \text{Vol}(\mathcal{E}_{tn,\eta}) = (1+\eta)^{tn/2}|\textbf{Q}^T\textbf{Q}|^{-1/2}\text{Vol}(\mathcal{B}(\sqrt{tn})),
\end{equation}
where $\text{Vol}(\mathcal{B}(\sqrt{tn}))$ is the volume of a $tn$-dimensional sphere of radius $\sqrt{tn}$. For brevity, we let $Pr(\bar{\mathcal{A}}) = (1+\delta)\frac{\text{Vol}(\mathcal{E}_{n,\eta})}{\text{Vol}(\mathcal{V}_F)}$. From (\ref{eq_message_rate_MIMO}), we have 
\begin{equation}
    \label{eq_vol_F}
    \text{Vol}(\mathcal{V}_F) = \text{Vol}(\mathcal{V}_l^C)\cdot2^{-n r_l} = \beta_l^{tn}\text{Vol}(\mathcal{V}_C)\cdot2^{-n r_l}.
\end{equation}
By combining (\ref{eq_vol_decision_region}) and (\ref{eq_vol_F}), we have
\begin{equation}
    \label{eq_ambiguity_prob_rate}
    Pr(\bar{\mathcal{A}}) \leq (1+\delta)\cdot 2^{-n\left[\frac{1}{2n}\log\frac{\beta_l^{2tn}}{|\boldsymbol{\Sigma}_1|}-r_l-\eta'\right]}
\end{equation}
where $\eta' = \frac{t}{2}\log(1+\eta)+\frac{1}{n}\log \frac{\text{Vol}(\mathcal{B}(\sqrt{tn}))}{\text{Vol}(\mathcal{V}_C)}$. From the fact that the constructed coarse lattices are good for covering, we have $G(\Lambda_C)\rightarrow \frac{1}{2\pi e}$ as $n\rightarrow\infty$, where $G(\Lambda_C)$ is the normalized second-order moment of $\Lambda_C$. Given $\text{Vol}(\mathcal{V}_C) = \left(\frac{\sigma^2(\Lambda_C)}{G(\Lambda_C)}\right)^{tn/2}$, we then have $\text{Vol}(\mathcal{V}_C) \rightarrow (2\pi e)^{tn/2}$ when $n\rightarrow\infty$.
By using the fact
\begin{equation}
    \label{eq_vol_ball_inf}
    \text{Vol}(\mathcal{B}(\sqrt{tn})) \rightarrow \frac{(2\pi e)^ {tn/2}}{\sqrt{tn\pi}} \text{ as } n\rightarrow\infty, 
\end{equation}
we can derive
\begin{equation}
    \label{eq_vol_fraction_inf}
    \frac{1}{n}\log \frac{\text{Vol}(\mathcal{B}(\sqrt{tn}))}{\text{Vol}(\mathcal{V}_C)} \rightarrow -\log (\sqrt{tn\pi})^{1/n} \rightarrow 0.
\end{equation}
The second right arrow comes from the fact that $\lim_{x\rightarrow\infty}(x)^{\lambda/x}=1,\forall\lambda>0$ when $x = tn\pi$ and $\lambda = \frac{t\pi}{2}$. Since $\eta>0$ is arbitrary, we can make $\eta'$ arbitrarily small with large enough $n$. Therefore, we can conclude that for arbitrary $\epsilon_1 >0$, $Pr(\bar{\mathcal{A}})\leq \epsilon_1/2$ for sufficiently large $n$ given that 
\begin{equation}
    \label{eq_rate_ambiguity}
    r_l < \frac{1}{2n}\log\frac{\beta_l^{2tn}}{|\boldsymbol{\Sigma}_1|}.
\end{equation}
The proof of (\ref{eq_achievable_rate_general}) will then be complete if we can show that $Pr(\bar{\textbf{z}}_1 \notin \mathcal{E}_{tn,\eta})\leq \epsilon_2/2$ for arbitrary $\eta, \epsilon_2>0$. Recall that 
\begin{equation}
    \label{eq_wrong_decode_error}
    Pr(\bar{\textbf{z}}_1 \notin \mathcal{E}_{tn,\eta}) = Pr(|\textbf{Q}\bar{\textbf{z}}_1|^2 > tn(1+\eta)),
\end{equation}
where $\bar{\textbf{z}}_1 = \textbf{W}\textbf{z} + \sum_l (\tilde a_l\textbf{I}_{tn}-\textbf{W}\bar{\textbf{H}}_l\bar{\textbf{B}}_l)(-\textbf{c}_l)$ is the effective noise with $\textbf{z}\sim \mathcal{N}(\textbf{0},\textbf{I}_{rn})$ and $\textbf{c}_l\sim \text{Uniform}(\mathcal{V}_C)$ for $l=1,2$. Note that $\textbf{z}, \textbf{c}_1, \textbf{c}_2$ are statistically independent by definition. To upper-bound this probability, we will consider a ``noisier" system with higher noise variance. We first add a Gaussian vector $\textbf{e}_3\sim \mathcal{N}(\textbf{0},(\sigma^2-1)\textbf{I}_{rn})$ to $\textbf{z}$ to make it noisier, where 
\begin{equation}
    \sigma^2 = \frac{r_{c}(\Lambda_C)^2}{tn}.
\end{equation}
The term $r_{c}(\Lambda_C)$ is the covering radius of $\Lambda_C$. From \cite{Gamal04}, we know $\sigma^2>1$. As $\sigma^2-1>0$, the noise $\textbf{e}_3$ is well-defined. We then replace $\textbf{c}_l$ with Gaussian vector $\textbf{e}_l\sim\mathcal{N}(\textbf{0},\sigma^2\textbf{I}_{tn})$. Recall $\sigma^2(\Lambda_C) =1$, thus $\textbf{e}_l$ has a larger variance. The considered noise is then replaced by
\begin{equation}
    \label{eq_noisier_noise}
    \bar{\textbf{z}}_1' = \textbf{W}(\textbf{z}+\textbf{e}_3) + \sum_l (\tilde a_l\textbf{I}_{tn}-\textbf{W}\bar{\textbf{H}}_l\bar{\textbf{B}}_l)\textbf{e}_l.
\end{equation}
We can then find that $\bar{\textbf{z}}_1'\sim \mathcal{N}(\textbf{0},\sigma^2\boldsymbol{\Sigma}_1)$. Let $f_{\textbf{c}_l}(\cdot)$ and $f_{\textbf{e}_l}(\cdot)$ denote the PDF of $\textbf{c}_l$ and $\textbf{e}_l$, respectively. Following the argument of \cite[Lemma 11]{Erez04} and \cite{Gamal04}, we can obtain
\begin{equation}
    \label{eq_pdf_upper_bound}
    f_{\textbf{c}_l}(\textbf{z}) \leq \left(\frac{r_{c}(\Lambda_C)}{r_{e}(\Lambda_C)}\right)^{tn} \exp(o(tn)) f_{\textbf{e}_l}(\textbf{z}),
\end{equation}
where $r_{e}(\Lambda_C)$ is the effective radius of $\Lambda_C$. When $n\rightarrow\infty$, $r_{c}(\Lambda_C)/r_{e}(\Lambda_C)\rightarrow 1$ from the fact that the constructed coarse lattices are good for covering. From the noisier construction, (\ref{eq_wrong_decode_error}) can be upper-bounded by
\begin{equation}
    \label{eq_wrong_decode_error_upper_bound}
    \begin{split}
        Pr(|\textbf{Q}\bar{\textbf{z}}_1|^2 > tn(1+\eta)) \leq \left[\left(\frac{r_{c}(\Lambda_C)}{r_{e}(\Lambda_C)}\right)^{tn} \exp(o(tn))\right]^2 Pr(|\textbf{Q}\bar{\textbf{z}}_1'|^2 \geq tn(1+\eta)).
    \end{split}
\end{equation}
Note that $\textbf{Q}\bar{\textbf{z}}_1'\sim \mathcal{N}(\textbf{0},\sigma^2\textbf{I}_{tn})$ by the construction of $\bar{\textbf{z}}_1'$ in (\ref{eq_noisier_noise}). Thus, $|\textbf{Q}\bar{\textbf{z}}_1'/\sigma|^2\sim \chi^2(tn)$. We can use the Chernoff bounding approach in \cite{Gamal04} to upper-bound $Pr(|\textbf{Q}\bar{\textbf{z}}_1'|^2 > tn(1+\eta))$, which gives
\begin{equation}
    \label{eq_wrong_decode_error_chernoff_bound}
    Pr(|\textbf{Q}\bar{\textbf{z}}_1'|^2 \geq tn(1+\eta)) \leq \min_{\alpha > 0}\; e^{-\alpha tn(1+\eta)} \mathbb{E}[e^{\alpha|\textbf{Q}\bar{\textbf{z}}_1'|^2}],
\end{equation}
where $\mathbb{E}[e^{\alpha|\textbf{Q}\bar{\textbf{z}}_1'|^2}] = \exp(-\frac{tn}{2}\ln(1-2\alpha \sigma^2))$. Therefore,
\begin{equation}
    \label{eq_wrong_decode_error_chernoff_bound_min}
    \begin{split}
        Pr(|\textbf{Q}\bar{\textbf{z}}_1'|^2 \geq tn(1+\eta)) & \leq \min_{\alpha > 0}\; e^{-\frac{tn}{2}[2\alpha (1+\eta)+\ln(1-2\alpha \sigma^2))]}\\
        & = \exp\left(-\frac{tn}{2}(\zeta - \ln \zeta - 1)\right),
    \end{split}
\end{equation}
where $\zeta = \frac{1+\eta}{\sigma^2}$. It is worth noting that $\sigma^2\rightarrow 1$ when $n\rightarrow\infty$. For arbitrary $\eta >0$, we have $\zeta > 1$ which leads to $\zeta - \ln \zeta - 1 >0$. We can then conclude that $Pr(\bar{\textbf{z}}_1 \notin \mathcal{E}_{tn,\eta})\leq \epsilon_2/2$ for arbitrary $\eta, \epsilon_2>0$ and sufficiently large $n$. Now we complete the proof of (\ref{eq_achievable_rate_general}).

To find the minimum $|\boldsymbol{\Sigma}_1|$, we rewrite $\boldsymbol{\Sigma}_1(\textbf{W})$ as
\begin{equation}
    \label{eq_Sigma_1}
    \begin{split}
        \boldsymbol{\Sigma}_1(\textbf{W}) = & \sum_{l}\tilde a_l^2\textbf{I}_{tn} +  \textbf{W}\left(\textbf{I}_{rn} + \sum_{l} \bar{\textbf{H}}_l\bar{\textbf{K}}_l\bar{\textbf{H}}_l^T\right)\textbf{W}^T - \left(\sum_{l}\tilde a_l \bar{\textbf{B}}_l^T \bar{\textbf{H}}_l^T\right) \textbf{W}^T - \textbf{W} \left(\sum_{l}\tilde a_l \bar{\textbf{H}}_l \bar{\textbf{B}}_l \right).
    \end{split}
\end{equation}
We can minimize $|\boldsymbol{\Sigma}_1(\textbf{W})|$ over $\textbf{W}$ by ``completing the square'', which gives
\begin{equation}
    \label{eq_optimal_W}
    \textbf{W}^* = \left(\sum_{l}\tilde a_l \bar{\textbf{B}}_l^T \bar{\textbf{H}}_l^T\right)\left(\textbf{I}_{rn} + \sum_{l} \bar{\textbf{H}}_l \bar{\textbf{K}}_l \bar{\textbf{H}}_l^T\right)^{-1}.
\end{equation}
Thus, 
\begin{equation}
    \label{eq_optimal_Sigma_1}
    \begin{split}
        \boldsymbol{\Sigma}_1(\textbf{W}^*) = &\sum_{l}\tilde a_l^2\textbf{I}_{tn} - \left(\sum_{l}\tilde a_l \bar{\textbf{B}}_l^T \bar{\textbf{H}}_l^T\right)\left(\textbf{I}_{rn} + \sum_{l} \bar{\textbf{H}}_l \bar{\textbf{K}}_l \bar{\textbf{H}}_l^T\right)^{-1} \left(\sum_{l}\tilde a_l \bar{\textbf{H}}_l \bar{\textbf{B}}_l \right).
    \end{split}
\end{equation}
To calculate $|\boldsymbol{\Sigma}_1(\textbf{W}^*)|$, we will use the matrix determinant lemma 
\begin{equation}\label{eq_matrix_determinant_lemma}
    |\textbf{D}-\textbf{C}\textbf{A}^{-1}\textbf{B}| = \frac{\left|\begin{array}{cc}
        \textbf{A} & \textbf{B} \\
        \textbf{C} & \textbf{D}
    \end{array}\right|}{|\textbf{A}|} = \frac{|\textbf{D}||\textbf{A}-\textbf{B}\textbf{D}^{-1}\textbf{C}|}{|\textbf{A}|},
\end{equation}
and Sylvester's determinant theorem: For a $m\times n$ matrix $\textbf{A}$ and a $n\times m$ matrix $\textbf{B}$, it holds that
\begin{equation}\label{eq_Sylvesters_determinant_theorem}
    |\textbf{I}_m+\textbf{A}\textbf{B}| = |\textbf{I}_n+\textbf{B}\textbf{A}|.
\end{equation}
With the help of (\ref{eq_matrix_determinant_lemma}) and (\ref{eq_Sylvesters_determinant_theorem}), we can simplify $|\boldsymbol{\Sigma}_1(\textbf{W}^*)|$ as
\begin{equation}
    \label{eq_optimal_det_Sigma_1}
    |\boldsymbol{\Sigma}_1(\textbf{W}^*)| = \frac{\left|\bar{\textbf{M}} \right|}{\left|\textbf{I}_{rn} + \sum_{l} \bar{\textbf{H}}_l \bar{\textbf{K}}_l\bar{\textbf{H}}_l^T \right|}, 
\end{equation}
where 
\begin{equation}
    \bar{\textbf{M}} = (\tilde a_1^2 + \tilde a_2^2)\textbf{I}_{tn} + (\tilde a_1 \bar{\textbf{B}}_2^T \bar{\textbf{H}}_2^T - \tilde a_2 \bar{\textbf{B}}_1^T \bar{\textbf{H}}_1^T)(\tilde a_1 \bar{\textbf{H}}_2 \bar{\textbf{B}}_2 - \tilde a_2 \bar{\textbf{H}}_1 \bar{\textbf{B}}_1 ).
\end{equation}
Therefore, the achievable rate to decode the first linear combination with coefficients $\textbf{a} = (a_1,a_2)^T$ is given by
\begin{equation}
    r_{l}^{(1)} = \frac{1}{2n} \log^+ 
    \frac{\beta_l^{2tn} \left|\textbf{I}_{rn} + \sum_{l} \bar{\textbf{H}}_l \bar{\textbf{K}}_l\bar{\textbf{H}}_l^T\right|}
    {\left|\bar{\textbf{M}}\right|}.
\end{equation}
Since $\textbf{I}_{rn} + \sum_{l} \bar{\textbf{H}}_l \bar{\textbf{K}}_l\bar{\textbf{H}}_l^T$ and $\bar{\textbf{M}}$ are both block diagonal matrices, $r_{l}^{(1)}$ can be written as
\begin{equation}
    \label{eq_achievable_rate_MIMO_1_sim}
    r_{l}^{(1)} = \frac{1}{2} \log^+ 
    \frac{\beta_l^{2t} \left|\textbf{I}_{r} + \sum_{l} \textbf{H}_l \textbf{K}_l\textbf{H}_l^T\right|}
    {\left|\textbf{M}\right|},
\end{equation}
where
\begin{equation}
    \textbf{M} = (\tilde a_1^2 + \tilde a_2^2)\textbf{I}_{t} + (\tilde a_1 \textbf{B}_2^T \textbf{H}_2^T - \tilde a_2 \textbf{B}_1^T \textbf{H}_1^T)(\tilde a_1 \textbf{H}_2 \textbf{B}_2 - \tilde a_2 \textbf{H}_1 \textbf{B}_1).
\end{equation}

Assume the first linear combination is decoded successfully, we can reconstruct $\sum_{l} \tilde a_l \textbf{c}_l = \sum_{l} a_l \tilde{\textbf{t}}_l + \sum_{l} a_l \beta_l \textbf{d}_l$. Let $\textbf{b} = (b_1,b_2)^T$ denote the computation coefficients of the second linear combination, such that the matrix $(\textbf{a}, \textbf{b})$ has full rank. To decode the second linear combination, the receiver computes
\begin{align*}
    \textbf{y}_2^\prime = & \textbf{F}\textbf{y} + \textbf{L} \sum_{l} a_l\beta_l \textbf{c}_l - \sum_{l} b_l \beta_l \textbf{d}_l \\
    = & \sum_l \textbf{F}\bar{\textbf{H}}_l \bar{\textbf{B}}_l \textbf{c}_l + \textbf{F}\textbf{z} + \textbf{L} \sum_{l} a_l\beta_l \textbf{c}_l - \sum_l b_l\beta_l \textbf{d}_l - \sum_l b_l\beta_l \textbf{c}_l + \sum_l b_l\beta_l \textbf{c}_l \\
    = & \underbrace{\textbf{F}\bar{\textbf{z}} + \sum_l (\tilde b_l \textbf{I}_{tn} - \textbf{F}\bar{\textbf{H}}_l \bar{\textbf{B}}_l - \tilde a_l \textbf{L})(-\textbf{c}_l) }_{\bar{\textbf{z}}_2} + \sum_l b_l \tilde{\textbf{t}}.
\end{align*}
Similar to decoding the first linear combination, the achievable rate to decode $\sum_l b_l \tilde{\textbf{t}}_l$ is given by
\begin{equation}
    r_{l}^{(2)} = \max_{\textbf{F},\textbf{L}}\quad \frac{1}{2n} \log^+ \frac{|\beta_l^2\textbf{I}_{tn}|}{|\boldsymbol{\Sigma}_2(\textbf{F},\textbf{L})|},
\end{equation}
where
\begin{equation}
    \label{eq_Sigma_2}
    \begin{split}
    \boldsymbol{\Sigma}_2(\textbf{F},\textbf{L}) = & \sum_{l}\tilde b_l^2 \textbf{I}_{tn} +  \textbf{F}\left(\textbf{I}_{rn} + \sum_{l} \bar{\textbf{H}}_l\bar{\textbf{K}}_l\bar{\textbf{H}}_l^T\right)\textbf{F}^T - \left(\sum_{l}\tilde b_l \bar{\textbf{B}}_l^T\bar{\textbf{H}}_l^T\right) \textbf{F}^T - \textbf{F} \left(\sum_{l}\tilde b_l \bar{\textbf{H}}_l \bar{\textbf{B}}_l \right)\\
    & + \sum_l \tilde a_l^2\textbf{L}\textbf{L}^T - \left(\sum_l \tilde a_l (\tilde b_l \textbf{I}_{tn} - \textbf{F}\bar{\textbf{H}}_l \bar{\textbf{B}}_l)\right) \textbf{L}^T - \textbf{L} \left(\sum_l \tilde a_l  (\tilde b_l \textbf{I}_{tn} - \bar{\textbf{B}}_l^T  \bar{\textbf{H}}_l^T \textbf{F}^T)\right).
    \end{split}
\end{equation}
Optimizing it over $\textbf{L}$ gives
\begin{equation}
    \label{eq_optimal_L}
    \textbf{L}^* = \left(\sum_{l}\tilde a_l^2\right)^{-1}\left(\sum_l \tilde a_l (\tilde b_l \textbf{I}_{tn} - \textbf{F}\bar{\textbf{H}}_l \bar{\textbf{B}}_l)\right).
\end{equation}
Plugging this into (\ref{eq_Sigma_2}) and simplifying it results in
\begin{equation}
    \label{eq_sigma_2_optimal_L}
    \begin{split} 
    \boldsymbol{\Sigma}_2(\textbf{F},\textbf{L}^*) = & \frac{(\tilde a_1 \tilde b_2 - \tilde a_2 \tilde b_1)^2}{\sum_{l}\tilde a_l^2}\textbf{I}_{tn} + \textbf{F} \textbf{M}_k \textbf{F}^T - \frac{\tilde a_1 \tilde b_2 - \tilde a_2 \tilde b_1}{\sum_{l}\tilde a_l^2}
    [(\tilde a_1 \bar{\textbf{B}}_2^T\bar{\textbf{H}}_2^T - \tilde a_2 \bar{\textbf{B}}_1^T\bar{\textbf{H}}_1^T) \textbf{F}^T \\
    & + \textbf{F} (\tilde a_1 \bar{\textbf{H}}_2\bar{\textbf{B}}_2 - \tilde a_2 \bar{\textbf{H}}_1 \bar{\textbf{B}}_1)],
    \end{split}
\end{equation}
where 
\begin{equation}
    \label{eq_Mk}
    \begin{split}
        \textbf{M}_k = & \textbf{I}_{rn} + \left(\sum_{l}\tilde a_l^2\right)^{-1}(\tilde a_1 \bar{\textbf{H}}_2 \bar{\textbf{B}}_2 - \tilde a_2 \bar{\textbf{H}}_1 \bar{\textbf{B}}_1 ) \cdot (\tilde a_1 \bar{\textbf{B}}_2^T \bar{\textbf{H}}_2^T - \tilde a_2 \bar{\textbf{B}}_1^T \bar{\textbf{H}}_1^T).
    \end{split}
\end{equation}
Optimizing (\ref{eq_sigma_2_optimal_L}) over $\textbf{F}$ gives
\begin{equation}
    \label{eq_optimal_F}
    \textbf{F}^* = \frac{\tilde a_1 \tilde b_2 - \tilde a_2 \tilde b_1}{\sum_{l}\tilde a_l^2}
    (\tilde a_1 \bar{\textbf{B}}_2^T\bar{\textbf{H}}_2^T -\bar{\textbf{B}}_1^T\bar{\textbf{H}}_1^T)
    \textbf{M}_k^{-1}.
\end{equation}
Thus,
\begin{equation}
    \label{eq_optimal_Sigma_2}
    \begin{split}
        \boldsymbol{\Sigma}_2(\textbf{F}^*,\textbf{L}^*) = &\frac{(\tilde a_1 \tilde b_2 - \tilde a_2 \tilde b_1)^2}{\sum_{l}\tilde a_l^2}\textbf{I}_{tn} - \left(\frac{\tilde a_1 \tilde b_2 - \tilde a_2 \tilde b_1}{\sum_{l}\tilde a_l^2} \right)^2 \\
        & \cdot (\tilde a_1 \bar{\textbf{B}}_2^T\bar{\textbf{H}}_2^T - \bar{\textbf{B}}_1^T\bar{\textbf{H}}_1^T) \textbf{M}_k^{-1} (\tilde a_1 \bar{\textbf{H}}_2\bar{\textbf{B}}_2 - \tilde a_2 \bar{\textbf{H}}_1 \bar{\textbf{B}}_1 ).
    \end{split}
\end{equation}
To calculate the determinant, we apply the matrix determinant lemma (\ref{eq_matrix_determinant_lemma}), which leads to
\begin{align}
    |\boldsymbol{\Sigma}_2(\textbf{F}^*,\textbf{L}^*)|
    = & \frac{(\tilde a_1 \tilde b_2 - \tilde a_2 \tilde b_1)^{2tn}}
    {\left|\bar{\textbf{M}}\right|} \label{eq_optimal_det_Sigma_2}
\end{align}
Therefore, the achievable rate to decode the second linear combination with coefficients $\textbf{b} = (b_1,b_2)^T$ is
\begin{equation}
    r_{l}^{(2)} = \frac{1}{2n} \log^+ 
    \frac{\beta_l^{2tn} {\left|\Bar{\textbf{M}}\right|}}
    {(\tilde a_1 \tilde b_2 - \tilde a_2 \tilde b_1)^{2tn}}.
\end{equation}
Since $\bar{\textbf{M}}$ is block diagonal, $r_{l}^{(2)}$ can be written as
\begin{equation}
    \label{eq_achievable_rate_MIMO_2_sim}
    r_{l}^{(2)} = \frac{1}{2} \log^+ 
    \frac{\beta_l^{2t} {\left|\textbf{M}\right|}}
    {(\tilde a_1 \tilde b_2 - \tilde a_2 \tilde b_1)^{2t}}.
\end{equation}

To have both linear combinations decoded successfully, the transmission rate of each user should not exceed the smaller achievable rate of both decoding processes. If $a_l$ or $b_l$ equals zero, which means one linear combination contains no information of user $l$, the achievable rate of user $l$ simply comes from decoding the other linear combination. Therefore, the proof of Theorem~\ref{thm_rate_pair} is complete.
\end{IEEEproof}

\begin{remark}
    when $t=r=1$, the results of Theorem~\ref{thm_rate_pair} reduce to Theorem~$2$ in \cite{Zhu17}.
\end{remark}

\subsection{Sum capacity achievability with CFMA-SCS}\label{sec_sum_capacity_achievability_SCS}
The sum capacity of a two-user MIMO Gaussian MAC with channel matrices $\textbf{H}_1,\textbf{H}_2$ is known to be \cite{gamal_kim_2011}
\begin{equation}
    \label{eq_sum_capacity}
    C_{sum} = \max_{\textbf{K}_1,\textbf{K}_2} \;\frac{1}{2}\log \left|\textbf{I}_{r} + \textbf{H}_1 \textbf{K}_1\textbf{H}_1^T + \textbf{H}_2 \textbf{K}_2\textbf{H}_2^T\right|,
\end{equation}
where the optimized input covariance matrices $\textbf{K}_1^*, \textbf{K}_2^*$ can be found by the iterative water-filling algorithm \cite{gamal_kim_2011}. Our derived results in Theorem~\ref{thm_rate_pair} hold for any input covariance matrices hence including the optimal $\textbf{K}_1$ and $\textbf{K}_2$. To simplify the presentation in the rest of this paper, for given $\textbf{H}_1$ and $\textbf{H}_2$ we define 
\begin{equation}
    \label{eq_c_sum_wo_log}
    C_d = \left|\textbf{I}_{r} + \textbf{H}_1 \textbf{K}_1^*\textbf{H}_1^T + \textbf{H}_2 \textbf{K}_2^*\textbf{H}_2^T\right|.
\end{equation}
We will then give the conditions when the sum capacity can be achieved by the CFMA scheme.

\begin{thm}\label{thm_sum_capacity}
    For a two-user MIMO Gaussian MAC given the channel matrices $\textbf{H}_1$ and $\textbf{H}_2$, with CFMA-SCS, the sum capacity is achievable by choosing $\textbf{a} = (1,1)$, $\textbf{b} = (1,0)$ or $(0,1)$, and $\beta_1/\beta_2 = \gamma$, if there exists some square matrices $\textbf{B}_1^*,\textbf{B}_2^*$ and real positive $\gamma$ such that
    \begin{align}
        & \textbf{B}_l^* \textbf{B}_l^{*T} = \textbf{K}_l^*, l = 1,2, \label{ineq_beta_condition_sumrate_B}\\
        \text{and }& g(\gamma,\textbf{B}_1^*,\textbf{B}_2^*) = f(\gamma,\textbf{B}_1^*,\textbf{B}_2^*) - \gamma^{t} \sqrt{C_d} \leq 0, \label{ineq_beta_condition_sumrate}
    \end{align}
where $\textbf{K}_l^{*}$ are the optimal input covariance matrices and 
\begin{equation}\label{eq_f_gamma}
    f(\gamma,\textbf{B}_1^*,\textbf{B}_2^*) = \left|(\gamma^2+1)\textbf{I}_{t} + (\gamma \textbf{B}_2^{*T} \textbf{H}_2^T - \textbf{B}_1^{*T} \textbf{H}_1^T)(\gamma \textbf{H}_2 \textbf{B}_2^* - \textbf{H}_1 \textbf{B}_1^*)\right|.
\end{equation}
\end{thm}
\begin{IEEEproof}
    When we choose $\textbf{a} = (1,1)$ and $\textbf{b} = (1,0)$, the achievable rate of user $2$ is given by $R_2 = r_2(\textbf{a},\boldsymbol{\beta})$ according to Theorem~\ref{thm_rate_pair}. If (\ref{ineq_beta_condition_sumrate}) holds, with $\gamma = \beta_1/\beta_2$ we have $\left| \textbf{M} \right|^2\leq \beta_1^{2t}\beta_2^{2t}C_d$, where $\textbf{M}$ is given in Theorem~\ref{thm_rate_pair} with $\textbf{B}_l = \textbf{B}_l^{*}$ for $l=1,2$. It can be further inferred that $r_1(\textbf{b}|\textbf{a},\boldsymbol{\beta})\leq r_1(\textbf{a},\boldsymbol{\beta})$. Thus the achievable rate of user $1$ is given by $R_1 = r_1(\textbf{b}|\textbf{a},\boldsymbol{\beta})$. The achievable sum rate in this case can be written as
    \begin{equation}
        \label{eq_sum_rate_b10}
        R_{sum} = R_1+R_2 = C_{sum} - t\log |a_1b_2-a_2b_1|.
    \end{equation}
    Since $|a_1b_2-a_2b_1| = 1$ for the chosen $\textbf{a}$ and $\textbf{b}$, the sum rate achieves the sum capacity. Similarly, when $\textbf{a} = (1,1)$ and $\textbf{b} = (0,1)$, we have $R_1 = r_1(\textbf{a},\boldsymbol{\beta})$ and $R_2 = r_2(\textbf{b}|\textbf{a},\boldsymbol{\beta})$. The achievable sum rate is also given by (\ref{eq_sum_rate_b10}), thus the proof is complete.
\end{IEEEproof}

It is worth noting that the choice of the precoding matrices $\textbf{B}_1,\textbf{B}_2$ is not unique for givenn covariance matrices $\textbf{K}_1$ and $\textbf{K}_2$. It can be observed from Theorem~\ref{thm_rate_pair} that the achievable rate pair, in particular, the matrix $\textbf{M}$ depends on $\textbf{B}_1,\textbf{B}_2$. Correspondingly, the sum-capacity-achieving condition in Theorem~\ref{thm_sum_capacity} depends on the choice of $\textbf{B}_1^*,\textbf{B}_2^*$ given the optimal covariance matrices $\textbf{K}_1^*$ and $\textbf{K}_2^*$. Notice that for any given $\gamma$, choosing $\textbf{B}_l$ to be the Cholesky decomposition will always satisfy the first condition (\ref{ineq_beta_condition_sumrate_B}). However, to make the second condition (\ref{ineq_beta_condition_sumrate}) easier to satisfy, we could minimize $g(\gamma,\textbf{B}_1^*,\textbf{B}_2^*)$ with respect to $\textbf{B}_l^*$ for $l=1,2$. Namely, for any given $\gamma$ we could solve the optimization problem
\begin{equation}
    \label{eq_optimize_B}
    \begin{split}
        \min_{\textbf{B}_1^*,\textbf{B}_2^*} & \quad g(\gamma,\textbf{B}_1^*,\textbf{B}_2^*)\\
        s.t. & \quad \textbf{B}_l^* \textbf{B}_l^{*T} = \textbf{K}_l^*, l = 1,2 .
    \end{split}
\end{equation}
In the current paper we do not focus on this particular aspect of optimization. In Sec.~\ref{sec_eg_perm}, we will give examples where we optimize $\textbf{B}_l^*$ for $l=1,2$ in a restricted class, and show that this ``partial optimization" gives better performance than the conventional $\textbf{B}_l^*$ obtained by the Cholesky decomposition.

It should be pointed out that for given channel coefficients and power constraints, (\ref{ineq_beta_condition_sumrate}) may not be satisfied for any choice of $\gamma$ or $\textbf{B}_1^*,\textbf{B}_2^*$ as we will show in Section~\ref{sec_sim} with some numerical examples. Thus, the sum capacity of the Gaussian MIMO channel is not always achievable with CFMA-SCS. This is in contrast to the result \cite{Zhu17} in the SISO case, where it is shown that the sum capacity (in fact any rate pair on the entire dominant face) is always achievable when the signal-to-noise ratio is high enough. But also notice that this result is derived from a specific code construction, where a long lattice codebook (with codeword length $tn$) is constructed for each user. The transmitted codeword is "chopped" into $t$ parts and each piece is fed into one antenna for transmission (after precoding). It will be shown that the capacity achievability can be improved for some channel setups with CFMA-PCS in Sec.~\ref{sec_PCS}.

\begin{remark}
    For given channel states and channel input covariance matrices, $g(\gamma)$ in (\ref{ineq_beta_condition_sumrate}) is a polynomial of $\gamma$ with the highest order $2t$. Its leading coefficient is given by $|\textbf{I}_t+\textbf{B}_2^{*T} \textbf{H}_2^{T}\textbf{H}_2 \textbf{B}_2^*|$, which is always positive. Thus, (\ref{ineq_beta_condition_sumrate}) can be satisfied if and only if $g(\gamma)$ has real roots. There are several numerical ways to check if a polynomial has real roots, for example, Sturm's theorem \cite{Basu16AlgorithmAlgebra}.
\end{remark}

\subsection{Case study for CFMA-SCS}\label{sec_case}
Apart from the numerical methods to check if the polynomial of interest has real roots, we study some special cases and provide more specific capacity-achieving conditions for them.

\subsubsection{SIMO MAC}\label{sec_simo}
In this subsection, we consider the special case where each transmitter only has a single transmit antenna, i.e., $t = 1$, while the receiver has $r \geq 1$ antennas. The channel coefficients are now given by vectors $\textbf{h}_1$ and $\textbf{h}_2$ of length $r$, and $K_1 = K_2 = P$ are two scalars. In this case, $C_d$ in (\ref{eq_c_sum_wo_log}) is given by
\begin{equation}
    \label{eq_sumcapacity_simo}
    C_d = \left|\textbf{I}_{r} + (\textbf{h}_1 \textbf{h}_1^T + \textbf{h}_2 \textbf{h}_2^T) P\right|.
\end{equation}
Note that the matrix $\textbf{h}_1 \textbf{h}_1^T + \textbf{h}_2 \textbf{h}_2^T$ is semi-positive definite and can be decomposed as $\textbf{U} \boldsymbol{\Lambda} \textbf{U}^T$, where $\textbf{U}$ is unitary, i.e., $\textbf{U}^T\textbf{U} = \textbf{I}_r$ and $\boldsymbol{\Lambda}$ is diagonal whose diagonal entries are the eigenvalues of $\textbf{h}_1 \textbf{h}_1^T + \textbf{h}_2 \textbf{h}_2^T$. Due to its special structure, the rank of $\textbf{h}_1 \textbf{h}_1^T + \textbf{h}_2 \textbf{h}_2^T$ is smaller than or equal to $2$. So it has at most two positive eigenvalues defined as $\lambda_1, \lambda_2$, and the other eigenvalues (if they exist) will be zero. Specifically, when $\textbf{h}_1$ and $\textbf{h}_2$ are collinear, we have $\lambda_2 = 0$ and $\lambda_1 = ||\textbf{h}_1||^2+||\textbf{h}_2||^2$ by calculating the roots of the characteristic polynomial of $\textbf{h}_1 \textbf{h}_1^T + \textbf{h}_2 \textbf{h}_2^T$ \cite{Andrilli23LinearAlgebra}. Thus, from Sylvester’s determinant theorem \cite{Sylvester1851}, (\ref{eq_sumcapacity_simo}) can be written as
\begin{equation}\label{eq_sumcapacity_simo_eigen_collinear}
    C_d = \left|\textbf{I}_{r} +  P \boldsymbol{\Lambda}\right| = 1+\lambda_1 P.
\end{equation}
If $\textbf{h}_1$ and $\textbf{h}_2$ are not collinear, we have $\lambda_1, \lambda_2>0$. In this case,  
\begin{equation}
    \label{eq_sumcapacity_simo_eigen}
    C_d = (1+\lambda_1 P)(1+\lambda_2 P),
\end{equation}
By analysing the capacity-achieving condition (\ref{ineq_beta_condition_sumrate}), we have the following lemma.
\begin{lem}\label{lem_SIMO}
    For a two-user SIMO Gaussian MAC given the channel coefficients $\textbf{h}_1$ and $\textbf{h}_2$, the sum capacity is achievable with CFMA-SCS if $\Delta \geq 0$, where
    \begin{equation}
        \label{eq_disc_simo}
        \Delta = (\sqrt{C_d} + 2 P \textbf{h}_1^T \textbf{h}_2)^2 - 4(1+P||\textbf{h}_1||^2)(1+P||\textbf{h}_2||^2).
    \end{equation}
    In particular, $\gamma$ should be chosen from the interval
    \begin{equation}
        \label{range_gamma_simo}
        \left[\frac{\sqrt{C_d} + 2 P \textbf{h}_1^T \textbf{h}_2-\sqrt{\Delta}}{2(1+P||\textbf{h}_2||^2)} ,\frac{\sqrt{C_d} + 2 P \textbf{h}_1^T \textbf{h}_2 + \sqrt{\Delta}}{2(1+P||\textbf{h}_2||^2)}\right],
    \end{equation}
    where $C_d$ is given by (\ref{eq_sumcapacity_simo}).
\end{lem}
\begin{IEEEproof}
    When $t=1$, $f(\gamma)$ in Theorem~\ref{thm_sum_capacity} is given by
    \begin{equation}
        \label{eq_f_gamma_simo}
        f(\gamma) = (\gamma^2+1) + (\gamma \sqrt{P} \textbf{h}_2^T - \sqrt{P} \textbf{h}_1^T)(\gamma \textbf{h}_2 \sqrt{P} - \textbf{h}_1 \sqrt{P}).
    \end{equation}
    The LHS of (\ref{ineq_beta_condition_sumrate}) in Theorem~\ref{thm_sum_capacity} is given by
    \begin{equation}
        \label{eq_g_gamma_simo}
        g(\gamma) = (1+P||\textbf{h}_2||^2)\gamma^2 - (\sqrt{C_d} + 2 P \textbf{h}_1^T \textbf{h}_2)\gamma + (1+P||\textbf{h}_1||^2).
    \end{equation}
    Note that $g(\gamma)$ is a second-order polynomial of $\gamma$ with a positive leading coefficient. Its discriminant $\Delta$ is given by (\ref{eq_disc_simo}). When $\Delta \geq 0$, $g(\gamma)$ will have two real roots and (\ref{ineq_beta_condition_sumrate}) is satisfied if $\gamma$ is within the two roots. Notice that both roots given by (\ref{range_gamma_simo}) are positive since $\sqrt{C_d} + 2 P \textbf{h}_1^T \textbf{h}_2>0$. Otherwise, if $\sqrt{C_d} + 2 P \textbf{h}_1^T \textbf{h}_2\leq 0$, we have $(\sqrt{C_d} + 2 P \textbf{h}_1^T \textbf{h}_2)^2 < (2 P \textbf{h}_1^T \textbf{h}_2)^2 \leq 4P^2||\textbf{h}_1||^2||\textbf{h}_2||^2$. Thus, $\Delta$ will become negative, which conflicts with $\Delta \geq 0$. Therefore, the sum capacity is achievable in this case and the proof is complete.
\end{IEEEproof}

\begin{remark}
    If we further let $r=1$, the channel becomes the SISO MAC. The sum capacity-achieving condition in Lemma~\ref{lem_SIMO} specializes to Theorem 3 Case II) in  \cite{Zhu17}.
\end{remark}

The following corollary shows that it is possible to achieve the sum capacity when the transmission power is large enough with some channel conditions.
\begin{coro}\label{coro_simo}
    Consider a two-user SIMO Gaussian MAC with channel coefficients $\textbf{h}_1$ and $\textbf{h}_2$.
    \begin{enumerate}
        \item When $\textbf{h}_1$ and $\textbf{h}_2$ are collinear, the sum capacity is achievable with CFMA-SCS if 
        \begin{equation}\label{ineq_power_channel_condtion_simo_col}
            \frac{P \textbf{h}_1^T \textbf{h}_2}{\sqrt{1+P(||\textbf{h}_1||^2+||\textbf{h}_2||^2)}} \geq \frac{3}{4}
        \end{equation}
        and $\gamma$ is chosen from the interval in (\ref{range_gamma_simo}).
        
        \item When $\textbf{h}_1$ and $\textbf{h}_2$ are linearly independent, the sum capacity is achievable with CFMA if $P\geq P^*$ and
        \begin{equation}\label{ineq_channel_condtion_simo_noncol}
            (\sqrt{\lambda_1\lambda_2} + 2\textbf{h}_1^T \textbf{h}_2)^2 > 4||\textbf{h}_1||^2||\textbf{h}_2||^2,
        \end{equation}
        where $\gamma$ is chosen within (\ref{range_gamma_simo}) and $P^*$ is the largest root of $\Delta(P)$ with $\Delta(P)$ given by (\ref{eq_disc_simo}) as a polynomial of $P$.
    \end{enumerate}
\end{coro}
\begin{IEEEproof}
    When $\textbf{h}_1$ and $\textbf{h}_2$ are collinear, we have $(\textbf{h}_1^T \textbf{h}_2)^2 = ||\textbf{h}_1||^2||\textbf{h}_2||^2$ and $C_d = 1+P(||\textbf{h}_1||^2+||\textbf{h}_2||^2)$ given by (\ref{eq_sumcapacity_simo_eigen_collinear}). Therefore, $\Delta$ in (\ref{eq_disc_simo}) can be simplified as
    \begin{equation}\label{eq_disc_simo_col}
        \Delta = 4 P \textbf{h}_1^T \textbf{h}_2 \sqrt{C_d} - 3C_d.
    \end{equation}
    If (\ref{ineq_power_channel_condtion_simo_col}) holds, we have $\Delta\geq 0$. Thus, the sum capacity is achievable in this case according to Lemma~\ref{lem_SIMO}.
    
    When $\textbf{h}_1$ and $\textbf{h}_2$ are linearly independent, if (\ref{ineq_channel_condtion_simo_noncol}) holds and with (\ref{eq_sumcapacity_simo_eigen}), we have $\Delta(P)$ given by (\ref{eq_disc_simo}) becoming a 2nd-order polynomial of $P$ with positive leading coefficient. Thus, as $P$ becomes large enough, $\Delta(P)$ will eventually become non-negative. Since $\Delta(0) = -3$, the largest root of $\Delta(P)$, denoted by $P^*$, will always be positive. If $P\geq P^*$, we have $\Delta(P)\geq 0$. Thus, the sum capacity is achievable in this case according to Lemma~\ref{lem_SIMO}.
\end{IEEEproof}

\subsubsection{Diagonal $2$-by-$2$ MIMO MAC}\label{sec_2by2_diagonal}
In this subsection, we consider the special case where the channel gains $\textbf{H}_1$ and $\textbf{H}_2$ are diagonal. In this case, the diagonal input covariance matrices with optimized power splitting can achieve the sum capacity. When $t=r=2$, we can denote
\begin{equation}
\label{eq_h_k_diagonal}
    \textbf{H}_l = \left[\begin{array}{cc}
        h_{l1} &  \\
         & h_{l2}
    \end{array}\right],
    \textbf{K}_l^* = \left[\begin{array}{cc}
        k_{l1} &  \\
         & k_{l2}
    \end{array}\right],
\end{equation}
where $k_{l1}, k_{l2}$ are the optimal power splitting and $k_{l1} + k_{l2} = P$ for $l=1,2$. We define some auxiliary parameters $c_{l1},c_{l2}$ for $l = 1,2$ as
\begin{equation}
    \label{eq_c_lj}
    c_{l1} = h_{l1}\sqrt{\frac{k_{l1}}{P}},\quad c_{l2} = h_{l2}\sqrt{\frac{k_{l2}}{P}}.
\end{equation}
For the given finite $\textbf{H}_1$ and $\textbf{H}_2$, $c_{l1}$ and $c_{l2}$ always take finite values since they only contain the channel coefficient and power splitting coefficient. We give a sufficient condition that guarantees to achieve the sum capacity for this case in the following lemma.
\begin{lem}\label{lem_diagonal}
    For a $2$-by-$2$ diagonal MIMO Gaussian MAC with channel matrices $\textbf{H}_1$ and $\textbf{H}_2$, when the power $P$ is large enough, the sum capacity is achievable with CFMA-SCS if either of the following conditions is satisfied.
    \begin{enumerate}
        \item $k_{11},k_{21}>0, \gamma = c_{11}/c_{21}$ and
        \begin{equation}
            \label{ineq_condition_diagonal_1}
            \left(\frac{c_{22}}{c_{21}} - \frac{c_{12}}{c_{11}}\right)^2 < \sqrt{\frac{c_{12}^2+c_{22}^2}{c_{11}^2+c_{21}^2}};
        \end{equation}
        \item $k_{12},k_{22}>0, \gamma = c_{12}/c_{22}$ and
        \begin{equation}
            \label{ineq_condition_diagonal_2}
            \left(\frac{c_{21}}{c_{22}} - \frac{c_{11}}{c_{12}}\right)^2 < \sqrt{\frac{c_{11}^2+c_{21}^2}{c_{12}^2+c_{22}^2}}.
        \end{equation}
    \end{enumerate}
\end{lem}
\begin{IEEEproof}
    With $\textbf{H}_l$ and $\textbf{K}_l^*$ given by (\ref{eq_h_k_diagonal}), $f(\gamma)$ in (\ref{ineq_beta_condition_sumrate}) is given by
    \begin{equation}
        \label{eq_f_gamma_diagonal}
        \begin{split}
            f(\gamma)
            = & \prod_{i=1}^2 \left[\gamma^2+1+(\gamma c_{2i}-c_{1i})^2 P\right],
        \end{split}
    \end{equation}
    and (\ref{eq_c_sum_wo_log}) is given by
    \begin{equation}
        \label{eq_sum_capacity_diagonal}
        C_d = \prod_{i=1}^2 \left[1+h_{1i}^2k_{1i} + h_{2i}^2k_{2i}\right] = \prod_{i=1}^2 \left[1+(c_{1i}^2 + c_{2i}^2)P\right].
    \end{equation}
    The capacity-achieving condition (\ref{ineq_beta_condition_sumrate}) is equivalent to 
    \begin{equation}
        \label{eq_q_gamma}
        q(\gamma) = f(\gamma)^2 - \gamma^{2t} C_d \leq 0.
    \end{equation}
    Following conditions 1) in Lemma~\ref{lem_diagonal}, we have $\gamma c_{21} - c_{11} = 0$. With this condition, $f(\gamma)$ in (\ref{eq_f_gamma_diagonal}) can be written as 
    \begin{equation}
        \label{eq_f_gamma_diagonal_1}
        f(\gamma) = (\gamma^2+1)(\gamma^2+1+(\gamma c_{22}-c_{12})^2 P).
    \end{equation}
    Therefore, with this given $\gamma$, $q(\gamma)$ in (\ref{eq_q_gamma}) becomes a 2nd-order polynomial of $P$ denoted by $\tilde{q}(P)$, whose leading coefficient is given by 
    \begin{align*}
        & (\gamma^2+1)^2(\gamma c_{22}-c_{12})^4 - \gamma^4 (c_{11}^2 + c_{21}^2)(c_{12}^2 + c_{22}^2) \\
        = & \gamma^4 \left[(1+\frac{1}{\gamma^2})^2(\gamma c_{22}-c_{12})^4 - (c_{11}^2 + c_{21}^2)(c_{12}^2 + c_{22}^2)\right] \\
        = & \gamma^4 \left[(1+\frac{c_{21}^2}{c_{11}^2})^2(\frac{c_{11}c_{22}}{c_{21}}-c_{12})^4 - (c_{11}^2 + c_{21}^2)(c_{12}^2 + c_{22}^2)\right] \\
        = & \gamma^4 (c_{11}^2 + c_{21}^2)^2 \left[(\frac{c_{22}}{c_{21}}-\frac{c_{12}}{c_{11}})^4 - \frac{c_{12}^2 + c_{22}^2}{c_{11}^2 + c_{21}^2}\right]
    \end{align*}
    According to (\ref{ineq_condition_diagonal_1}), this leading coefficient is negative. Thus when $P$ is large enough, $\tilde{q}(P)$ will become negative, which satisfies (\ref{ineq_beta_condition_sumrate}) in Theorem~\ref{thm_sum_capacity} and the sum capacity is achievable. Condition 2) in Lemma~\ref{lem_diagonal} is symmetric to Condition 1) and can be proved through the same procedure. 
\end{IEEEproof}

It is worth noting that in both conditions of Lemma~\ref{lem_diagonal}, $\gamma$ is chosen as a constant. Therefore, $q(\gamma)$ in (\ref{eq_q_gamma}) can be regarded as a quadratic function of $P$ denoted by $\tilde{q}(P)$, with a negative leading coefficient. Since $\tilde{q}(0)>0$ and $\tilde{q}(\infty)<0$, this quadratic function always has a positive real root. We can lower bound $P$ by this root to achieve the sum capacity. However, since its expression is too complicated, we omit this lower bound in the lemma.

\begin{remark}
    If the optimal power splitting follows neither of the conditions in Lemma~\ref{lem_diagonal}, i.e., $k_{11} = k_{22} = 0$ or $k_{12}=k_{21} = 0$, the two-user $2$-by-$2$ MIMO MAC is equivalent to two independent point-to-point SISO channels. In this case, no multiple access technique is required.
\end{remark}

\subsubsection{Channels with a similar SVD structure}\label{sec_decomposition}
In this subsection, we consider the special case where $\textbf{H}_1\textbf{B}_1^*$ and $\textbf{H}_2\textbf{B}_2^*$ have special singular value decomposition (SVD). Recall that $\textbf{B}_l^*$ satisfies $\textbf{B}_l^*\textbf{B}_l^{*T} = \textbf{K}_l^*$ for $l = 1,2$. Since the SVD always exists for any matrix, we can let $\textbf{H}_1\textbf{B}_1^* = \textbf{S}_1\textbf{V}_1\textbf{D}_1^T$ and $\textbf{H}_2\textbf{B}_2^* = \textbf{S}_2\textbf{V}_2\textbf{D}_2^T$, where $\textbf{S}_l$ and $\textbf{D}_l$ are orthogonal from the definition of SVD, i.e., $\textbf{S}_l^T\textbf{S}_l = \textbf{D}_l^T\textbf{D}_l = \textbf{I}$ for $l=1,2$. We consider the special case where $\textbf{S}_1 = \textbf{S}_2$ and $\textbf{D}_1=\textbf{D}_2$. Let $k = \max(rank(\textbf{V}_1),rank(\textbf{V}_2))$. Then we can write $\textbf{V}_l = diag(\lambda_{l1},\ldots,\lambda_{lk})$ for $l=1,2$. We further assume $t=r=k$. The term $C_d$ in (\ref{eq_c_sum_wo_log}) will be given by 
\begin{align}
    C_d = & \left|\textbf{I}_r + \textbf{S}\textbf{V}_1\textbf{D}^T\textbf{D}\textbf{V}_1\textbf{S}^T + \textbf{S}\textbf{V}_2\textbf{D}^T\textbf{D}\textbf{V}_2\textbf{S}^T \right|\nonumber \\
    = & \left|\textbf{I}_k + \textbf{V}_1^2 + \textbf{V}_2^2 \right|\label{eq_c_d_svd_matrix}  \\
    = & \prod_{i=1}^k \left(1+\lambda_{1i}^2+\lambda_{2i}^2\right)\label{eq_c_d_svd},
\end{align} 
where (\ref{eq_c_d_svd_matrix}) is derived with the help of (\ref{eq_Sylvesters_determinant_theorem}). We can also rewrite $f(\gamma)$ in (\ref{eq_f_gamma}) as 
\begin{align}
    f(\gamma) = & \left|(\gamma^2+1)\textbf{I}_{t} + (\gamma \textbf{B}_2^{*T} \textbf{H}_2^T - \textbf{B}_1^{*T} \textbf{H}_1^T)(\gamma \textbf{H}_2 \textbf{B}_2^* - \textbf{H}_1 \textbf{B}_1^*)\right| \nonumber\\
    = & \left|(\gamma^2+1)\textbf{I}_{t} + (\gamma \textbf{D}\textbf{V}_2\textbf{S}^T - \textbf{D}\textbf{V}_1\textbf{S}^T)(\gamma \textbf{S}\textbf{V}_2\textbf{D}^T - \textbf{S}\textbf{V}_1\textbf{D}^T)\right|\nonumber\\
    = & \left|(\gamma^2+1)\textbf{I}_{k} + (\gamma \textbf{V}_2 - \textbf{V}_1)^2\right|\label{eq_f_gamma_svd_matrix}\\
    = & \prod_{i=1}^k \left[\gamma^2+1 + (\gamma \lambda_{2i} - \lambda_{1i})^2\right]\label{eq_f_gamma_svd}.
\end{align}

Based on the above analysis, we have the following lemma for the sum capacity achievability.
\begin{lem}\label{lem_sum_capacity_condition_svd}
    For a $k$-by-$k$ MIMO MAC with $\textbf{H}_1\textbf{B}_1^* = \textbf{S}\textbf{V}_1\textbf{D}^T$ and $\textbf{H}_2\textbf{B}_2^* = \textbf{S}\textbf{V}_2\textbf{D}^T$, where $k= \max(rank(\textbf{V}_1),rank(\textbf{V}_2))$, $\textbf{S},\textbf{D}$ are orthogonal and $\textbf{V}_l = diag(\lambda_{l1},\ldots,\lambda_{lk})$ for $l=1,2$, the sum capacity is achievable with CFMA if 
    \begin{equation}
        \label{eq_sum_capacity_achieving_svd}
        \prod_{i=1}^k \left[\left(1+\lambda_{2i}^2\right)\gamma^2 - \left(2\lambda_{1i}\lambda_{2i} + \sqrt{1+\lambda_{1i}^2+\lambda_{2i}^2}\right) \gamma + 1+\lambda_{1i}^2\right] \leq 0.
    \end{equation}
\end{lem}
\begin{IEEEproof}
    We plug (\ref{eq_c_d_svd}) and (\ref{eq_f_gamma_svd}) into the sum capacity achieving condition of (\ref{ineq_beta_condition_sumrate}), which gives 
    \begin{align}
    f(\gamma) - \gamma^{t} \sqrt{C_d} = & \prod_{i=1}^k \left[\gamma^2+1 + (\gamma \lambda_{2i} - \lambda_{1i})^2 - \gamma \sqrt{1+\lambda_{1i}^2+\lambda_{2i}^2}\right] \\
    = & \prod_{i=1}^k \left[\left(1+\lambda_{2i}^2\right)\gamma^2 - \left(2\lambda_{1i}\lambda_{2i} + \sqrt{1+\lambda_{1i}^2+\lambda_{2i}^2}\right) \gamma + 1+\lambda_{1i}^2\right] \leq 0.
\end{align}
\end{IEEEproof}

The LHS of (\ref{eq_sum_capacity_achieving_svd}) is a product of several quadratic functions of $\gamma$ with positive leading coefficients. It can be observed that if any of the functions of $\gamma$, say $g_i(\gamma)$ has real roots, we can always find some particular $\gamma$ that satisfies (\ref{eq_sum_capacity_achieving_svd}). An obvious example is to choose $\gamma$ as a real root of $g_i(\gamma)$, which makes the LHS of (\ref{eq_sum_capacity_achieving_svd}) equal zero. We conclude this observation as the following corollary.
\begin{coro}
    For a $k$-by-$k$ MIMO MAC with $\textbf{H}_1\textbf{B}_1^* = \textbf{S}\textbf{V}_1\textbf{D}^T$ and $\textbf{H}_2\textbf{B}_2^* = \textbf{S}\textbf{V}_2\textbf{D}^T$, where $k= \max(rank(\textbf{V}_1), rank(\textbf{V}_2))$, $\textbf{S},\textbf{D}$ are orthogonal and $\textbf{V}_l = diag(\lambda_{l1},\ldots,\lambda_{lk})$ for $l=1,2$, the sum capacity is achievable with CFMA-SCS if If there exists any $i\in\{1,\ldots,k\}$ such that
    \begin{equation}
        \label{eq_sum_capacity_achieving_svd_delta}
        4 \lambda_{1i}\lambda_{2i} - 3 \sqrt{1+\lambda_{1i}^2+\lambda_{2i}^2} \geq 0.
    \end{equation}
\end{coro}
\begin{IEEEproof}
    Let $g_i(\gamma)$ denote the quadratic function 
    \begin{equation*}
        \left(1+\lambda_{2i}^2\right)\gamma^2 - \left(2\lambda_{1i}\lambda_{2i} + \sqrt{1+\lambda_{1i}^2+\lambda_{2i}^2}\right) \gamma' + 1+\lambda_{1i}^2
    \end{equation*}
    With (\ref{eq_sum_capacity_achieving_svd_delta}), the discriminant of $g_i(\gamma)$ is given by
    \begin{align*}
        & \left(2\lambda_{1i}\lambda_{2i} + \sqrt{1+\lambda_{1i}^2+\lambda_{2i}^2}\right)^2 - 4\left(1+\lambda_{2i}^2\right)\left(1+\lambda_{1i}^2\right)\\
        = & 4 \lambda_{1i} \lambda_{2i} \sqrt{1+\lambda_{1i}^2+\lambda_{2i}^2} - 3 (1+\lambda_{1i}^2+\lambda_{2i}^2)\\
        = & \sqrt{1+\lambda_{1i}^2+\lambda_{2i}^2} \left(4 \lambda_{1i}\lambda_{2i} - 3 \sqrt{1+\lambda_{1i}^2+\lambda_{2i}^2}\right) \geq 0.
    \end{align*}
    Thus, $g_i(\gamma)$ has real roots. The sum capacity achieving condition (\ref{eq_sum_capacity_achieving_svd}) in Lemma~\ref{lem_sum_capacity_condition_svd} can be satisfied with properly chosen $\gamma$. 
\end{IEEEproof}

\begin{remark}\label{remark_decomposition}
    We can further observe that if there exists $i\in\{1,\ldots,k\}$ such that $\lambda_{1i},\lambda_{2i} > \sqrt{3/2}$, the sum capacity-achieving condition can be satisfied. We will give an example later in Sec.~\ref{sec_eg_individual_codebook}.
\end{remark}

\section{The CFMA parallel coding scheme (PCS)}\label{sec_PCS}
In this section, we consider a different coding scheme for the two-user MIMO MAC, where we construct an individual codebook with codeword length $n$ for each transmit antenna. We can write the system model in the classical time-space form, i.e.,
\begin{equation}
\label{eq_channel_output_independent_codebook}
    \textbf{Y} = \textbf{H}_1 \textbf{X}_1 + \textbf{H}_2 \textbf{X}_2 + \textbf{Z},
\end{equation}
where $\textbf{Y},\textbf{Z} \in \mathbb{R}^{r\times n}$, $\textbf{X}_l \in \mathbb{R}^{t\times n}$ for $l=1,2$ and $\textbf{H}_l\in\mathbb{R}^{r\times t}$ are given in Sec.~\ref{sec_model}. 

\subsection{Transmission scheme for CFMA-PCS}\label{sec_coding_PCS}
In CFMA-PCS, each user can have multiple independent codebooks. Depending on the rank of the input covariance matrices $\textbf{K}_1,\textbf{K}_2$, there may be some inactive transmit antennas. For example, the optimal input covariance matrices are derived from iterative water-filling thus it is possible for some transmit antenna to have zero power allocation, namely becoming inactive. We will not construct codebooks for those inactive transmit antennas. Assume both users have $t_1$ and $t_2$ active antennas, respectively, where $t_l = rank(\textbf{K}_l)$ for $l=1,2$. We construct lattice codebook $\mathcal{C}_i$ for $i=1,\ldots,t_1+t_2$ with the second moment of the coarse lattice given by $\beta_i^2$. Let $\textbf{t}_i\in\mathbb{R}^n$ stand for a codeword in codebook $\mathcal{C}_i$. We then construct the normalized and dithered codeword $\textbf{c}_i$ as
\begin{equation}
    \label{eq_c_i_independent_codebook}
    \textbf{c}_i = [\textbf{t}_i/\beta_i+\textbf{d}_i] \mod \Lambda_i^C/\beta_i \text{  for } i = 1,\ldots,t_1+t_2.
\end{equation}
The channel input will be given by
\begin{equation}
    \label{eq_channel_input_independent_codebook}
    \textbf{X}_1 = \textbf{B}_1\left[\begin{array}{c}
        \textbf{c}_1^T      \\
        \vdots \\
        \textbf{c}_{t_1}^T \\
        \textbf{0}_{(t-t_1)\times n}
    \end{array}\right], \;
    \textbf{X}_2 = \textbf{B}_2\left[\begin{array}{c}
        \textbf{c}_{t_1+1}^T      \\
        \vdots \\
        \textbf{c}_{t_1+t_2}^T \\
        \textbf{0}_{(t-t_2)\times n}
    \end{array}\right],
\end{equation}
where $\textbf{B}_l\in\mathbb{R}^{t\times t}$ are given by the Cholesky decomposition of $\textbf{K}_l$ such that $\textbf{B}_l\textbf{B}_l^T=\textbf{K}_l$ for $l=1,2$. Recall that $rank(\textbf{K}_l) = t_l$ for $l=1,2$. We can write $\textbf{H}_l\textbf{B}_l$ as
\begin{equation}\label{eq_HB}
    \textbf{H}_l\textbf{B}_l = [\tilde{\textbf{H}}_{l}\; \textbf{0}_{r\times (t-t_l)}],
\end{equation}
where $\tilde{\textbf{H}}_{l} \in\mathbb{R}^{r\times t_l}$ include the first $t_l$ columns of $\textbf{H}_l\textbf{B}_l$ for $l=1,2$. The channel output can be given by
\begin{equation}
    \label{eq_channel_output_simo_independent_codebook}
    \textbf{Y} = \tilde{\textbf{H}}\textbf{C} + \textbf{Z},
\end{equation}
where $\tilde{\textbf{H}} = \left[\tilde{\textbf{H}}_1\;\tilde{\textbf{H}}_2\right]\in\mathbb{R}^{r\times (t_1+t_2)}$ and $\textbf{C} = \left[\textbf{c}_1\;\cdots\;\textbf{c}_{t_1+t_2}\right]^T\in\mathbb{R}^{(t_1+t_2)\times n}$. Since $\textbf{c}_i$ are normalised codewords from individual codebooks, this model can be regarded as a $(t_1+t_2)$-user SIMO MAC with an effective channel matrix $\tilde{\textbf{H}}$. Note that similar models have been studied in some literature such as \cite{Zhu17},\cite{nazer16expand} and we will apply the same decoding process here. The receiver will successively decode $t_1+t_2$ linearly independent integer linear combinations to recover the individual messages. For linear combination $j$ with integer coefficients $\textbf{a}_j = (a_{1j},\ldots,a_{t_1+t_2,j})^T \in\mathbb{Z}^{t_1+t_2}$ and some equalization vector $\textbf{b}_j,\textbf{q}_{j-1}$, the receiver computes
\begin{align*}
    & \textbf{b}_j^T \textbf{Y} - \textbf{a}_j^T \textbf{E} \textbf{D} + \textbf{q}_{j-1}^T \textbf{A}_{j-1} \textbf{E}\textbf{C}  \\
    = & \textbf{a}_j^T \tilde{\textbf{T}} + \underbrace{(\textbf{a}_j^T\textbf{E} - \textbf{b}_j^T\tilde{\textbf{H}} - \textbf{q}_{j-1}^T \textbf{A}_{j-1} \textbf{E}) (-\textbf{C}) + \textbf{b}_j^T\textbf{Z}}_{\text{effective noise}},
\end{align*}
where $\textbf{E} = \text{diag}(\beta_1,\ldots,\beta_{t_1+t_2})$, $\textbf{D} = \left[\textbf{d}_1\cdots\textbf{d}_{t_1+t_2}\right]^T$, $\textbf{A}_j = [\textbf{a}_1\cdots\textbf{a}_j]^T$ and $\tilde{\textbf{T}} = \left[\tilde{\textbf{t}}_1\cdots\tilde{\textbf{t}}_{t_1+t_2}\right]^T$ with $\tilde{\textbf{t}}_i$ given in Sec.~\ref{sec_rate_SCS}. After successfully decoding $t_1+t_2$ linearly independent combinations, the receiver can recover all the messages of all the transmit antennas. By correctly combining the $t_1+t_2$ messages, the message of each user can be recovered.

\subsection{Achievable rates with CFMA-PCS}
According to \cite[Theorem~$4$]{Zhu17} and \cite[Theorem~$2$]{nazer16expand}, the following message rate is achievable for codebook $\mathcal{C}_i$,
\begin{equation}
    \label{eq_antenna_achievable_rate_independent_codebook}
    r_i < \min_{j:\;a_{ij}\neq 0} \left(\frac{1}{2}\log^+ \frac{\beta_i^2}{\sigma_j^2}\right),
\end{equation}
where $\sigma_j^2$ denotes the effective noise variance of the $j$-th linear combination and is given by
\begin{equation}
    \label{eq_sigma_j}
    \sigma_j^2 = ||\textbf{b}_j||^2 + ||\textbf{E}\textbf{a}_j - \tilde{\textbf{H}}^T\textbf{b}_j - \textbf{E}\textbf{A}_{j-1}^T\textbf{q}_{j-1} ||^2.
\end{equation}
Similar to \cite[Lemma~$6$]{nazer16expand}, we find that the minimum effective noise variance can be achieved by choosing the equalization vector $\textbf{b}_j,\textbf{q}_{j-1}$ as 
\begin{align}
    \hat{\textbf{b}}_j^T = & (\textbf{a}_j^T \textbf{E}-\textbf{q}_{j-1}^T\textbf{A}_{j-1}\textbf{E})\tilde{\textbf{H}}^T(\textbf{I}+\tilde{\textbf{H}}\tilde{\textbf{H}}^T)^{-1}, \label{eq_optimal_b_j}\\
    \hat{\textbf{q}}_{j-1} = & \textbf{a}_j^T\textbf{E}\textbf{L}^T\textbf{L}\textbf{E}\textbf{A}_{j-1}^T(\textbf{A}_{j-1}\textbf{E}\textbf{L}^T\textbf{L}\textbf{E}\textbf{A}_{j-1}^T)^{-1},\label{eq_optimal_q_j}
\end{align}
where $\textbf{L}$ satisfies
\begin{equation}
    \label{eq_LTL}
    \textbf{L}^T\textbf{L} = \textbf{I} -\tilde{\textbf{H}}^T(\textbf{I}+\tilde{\textbf{H}}\tilde{\textbf{H}}^T)^{-1}\tilde{\textbf{H}} = (\textbf{I}+\tilde{\textbf{H}}^T\tilde{\textbf{H}})^{-1}.
\end{equation}
The second equality in (\ref{eq_LTL}) holds from Woodbury Matrix Identity. The minimum effective noise variance can be then written as 
\begin{equation}
\label{eq_optimal_sigma_j_M}
    \hat\sigma_{j}^2 = ||\textbf{M}_{j-1}\textbf{L}\textbf{E}\textbf{a}_j||^2,
\end{equation}
where 
\begin{equation}
    \label{eq_M_j}
    \textbf{M}_{j-1} = \textbf{I}- \textbf{L}\textbf{E}\textbf{A}_{j-1}^T(\textbf{A}_{j-1}\textbf{E}\textbf{L}^T\textbf{L}\textbf{E}\textbf{A}_{j-1}^T)^{-1}\textbf{A}_{j-1}\textbf{E}\textbf{L}^T.
\end{equation}

\subsection{Sum capacity achievability with CFMA-PCS}\label{sec_sum_capacity_PCS}
Based on the achievable rate in (\ref{eq_antenna_achievable_rate_independent_codebook}) and the minimum effective noise variance in (\ref{eq_optimal_sigma_j_M}), we provide a sufficient condition for CFMA-PCS to achieve the sum capacity in the following theorem.
\begin{thm}
\label{thm_capacity_achieve_individual_codebook}
    For a two-user MIMO Gaussian MAC given the channel matrices $\textbf{H}_1$ and $\textbf{H}_2$, let $t_l=rank(\textbf{K}^*_l)$ for $l=1,2$, where $\textbf{K}^*_l$ are the optimal input covariance matrices. With CFMA-PCS, the sum capacity is achievable if there exits a unimodular matrix $\textbf{A}\in\mathbb{Z}^{t_1+t_2}$ ($|\textbf{A}| = 1$) and a permutation $\Pi$ of the set $\{1,\ldots,t_1+t_2\}$ such that $\beta_i^2 \geq \hat\sigma_{\Pi(i)}^2$ for $i = 1,\ldots,t_1+t_2$ and 
    \begin{equation}
    \label{capacity_achieve_individual_codebook}
    \hat\sigma_{\Pi(1)}^2 \leq \hat\sigma_{\Pi(2)}^2 \leq \cdots \leq \hat\sigma_{\Pi(t_1+t_2)}^2,
\end{equation}
where $\hat\sigma_{\Pi(i)}^2$ is given in (\ref{eq_optimal_sigma_j_M}) and $\tilde{\textbf{H}}$ in (\ref{eq_LTL}) is derived with $\textbf{B}_1^*,\textbf{B}_2^*$ given by the Cholesky decomposition of $\textbf{K}_1^*,\textbf{K}_2^*$, respectively.
\end{thm}
\begin{IEEEproof}
    With CFMA-PCS, given the conditions in Theorem~\ref{thm_capacity_achieve_individual_codebook}, the achievable message rate of codebook $\mathcal{C}_i$ is given by 
    \begin{equation}
    \label{eq_antenna_achievable_rate_independent_codebook_optimal}
    r_i = \frac{1}{2}\log \frac{\beta_i^2}{\hat\sigma_{\Pi(i)}^2}.
\end{equation}
The sum rate in this case is given by $R_{sum} = \sum_{i=1}^{t_1+t_2} \frac{1}{2} \log\frac{\beta_i^2}{\hat\sigma_{\Pi(i)}^2}$. Similar to \cite[Sec.IV]{Zhu17} and \cite[Lemma~$8$]{nazer16expand}, with unimodular $\textbf{A}$ we can find that
\begin{equation}\label{eq_r_sum_individual_codebook}
    R_{sum} = \frac{1}{2} \log |\textbf{I}_r + \tilde{\textbf{H}}\tilde{\textbf{H}}^T| = \frac{1}{2} \log |\textbf{I}_r + \tilde{\textbf{H}}_1\tilde{\textbf{H}}_1^T + \tilde{\textbf{H}}_2\tilde{\textbf{H}}_2^T|.
\end{equation}
From (\ref{eq_HB}), we have $\textbf{H}_l\textbf{B}_l^* = [\tilde{\textbf{H}}_{l}\; \textbf{0}_{r\times (t-t_l)}]$ for $l=1,2$. Therefore,
\begin{equation}\label{eq_HB_H_tilde}
    \tilde{\textbf{H}}_l\tilde{\textbf{H}}_l^T = \textbf{H}_l\textbf{B}_l^*\textbf{B}_l^{*T}\textbf{H}_l^T = \textbf{H}_l\textbf{K}_l^*\textbf{H}_l^T,\quad l=1,2.
\end{equation}
The sum rate can be rewritten as 
\begin{equation}\label{eq_r_sum_individual_codebook_c_sum}
    R_{sum} = \frac{1}{2}\log \left|\textbf{I}_{r} + \textbf{H}_1 \textbf{K}_1^*\textbf{H}_1^T + \textbf{H}_2 \textbf{K}_2^*\textbf{H}_2^T\right|,
\end{equation}
which is equivalent to the sum capacity defined in (\ref{eq_sum_capacity}).
\end{IEEEproof}
\begin{remark}
    Now we comment on the difference between CFMA-PCS and CFMA-SCS.
    In CFMA-PCS, we construct $2t$ codebooks with codeword length $n$ in the encoding phase. Recall that $t$ is the number of transmit antennas for each user. In the decoding phase, $2t$ linear combinations are required to recover all the messages. However, in CFMA-SCS, only two codebooks are required but with longer codeword length $tn$. Consequently, we only need to decode two linear combinations to recover individual messages. Intuitively, the main coding difference between CFMA-SCS and CFMA-PCS is which information is distribute to the multiple transmit antennas. In CFMA-SCS, we first encode the messages to the lattice codewords and the codewords are separated into $t$ pieces, where each piece will be distributed to a transmit antenna. However, in CFMA-PCS, we first separate and distribute the messages to each transmit antenna. Then the message piece will be encoded to lattice codewords individually to be transmitted. It is difficult to compare these two schemes analytically. In Sec.~\ref{sec_sim}, we compare the two schemes with several numerical examples and comment on their suitability.
\end{remark}

\begin{remark}
    When $t=1$ (SIMO MAC), it is easy to see that CFMA-SCS and CFMA-PCS are identical, while when $t>1$, CFMA-PCS requires at least four codebooks. The sum capacity achieving condition in (\ref{capacity_achieve_individual_codebook}) will be equivalent to at least three inequalities. It becomes difficult to analyse the explicit form of the condition for the special cases as we have done in Sec.~\ref{sec_case}. In this case, we will provide some numerical results for the generic MIMO case under CFMA-PCS in Sec.~\ref{sec_eg_individual_codebook}.
\end{remark}

\section{Numerical result}\label{sec_sim}

In this section, we will study some SIMO and $2$-by-$2$ MIMO examples to illustrate our findings in previous sections. In particular, we will first consider different channel setups and see whether CFMA-SCS can achieve the sum capacity with given decoding coefficients $\textbf{a}=(1,1), \textbf{b} = (1,0)$ or $(0,1)$. We will then look at the problem of optimizing $\textbf{B}_1^*,\textbf{B}_2^*$ discussed after Theorem~\ref{thm_sum_capacity}, where permutation encoding is shown to improve the performance in CFMA-SCS. We will finally compare the sum capacity achievability between CFMA-SCS and CFMA-PCS in different channel conditions.

\subsection{Sum capacity achievability with CFMA-SCS}\label{sec_eg}
To intuitively understand whether CFMA-SCS can achieve the sum capacity under different channel conditions, we conduct numerical studies with different channel models, including SIMO MAC (analyzed in Sec.~\ref{sec_simo}), MAC with diagonal MIMO channels (analyzed in Sec.~\ref{sec_2by2_diagonal}), and MAC with generic MIMO channels. In our simulation, the channel coefficients for each realization are randomly generated from a uniform distribution. We check whether CFMA-SCS can achieve the sum capacity by (\ref{ineq_beta_condition_sumrate}) with any $\gamma$. In each case, we generate $10^3$ realizations and count the number when the sum capacity is achievable with CFMA (i.e., if (\ref{ineq_beta_condition_sumrate}) is satisfied) for different power constraints. Then we can calculate the ratio of this number to the total number of channel realizations. We will denote this ratio as $R_A$ in the later discussion. 

\begin{figure}
    \centering
    \includegraphics[width = 0.85\linewidth]{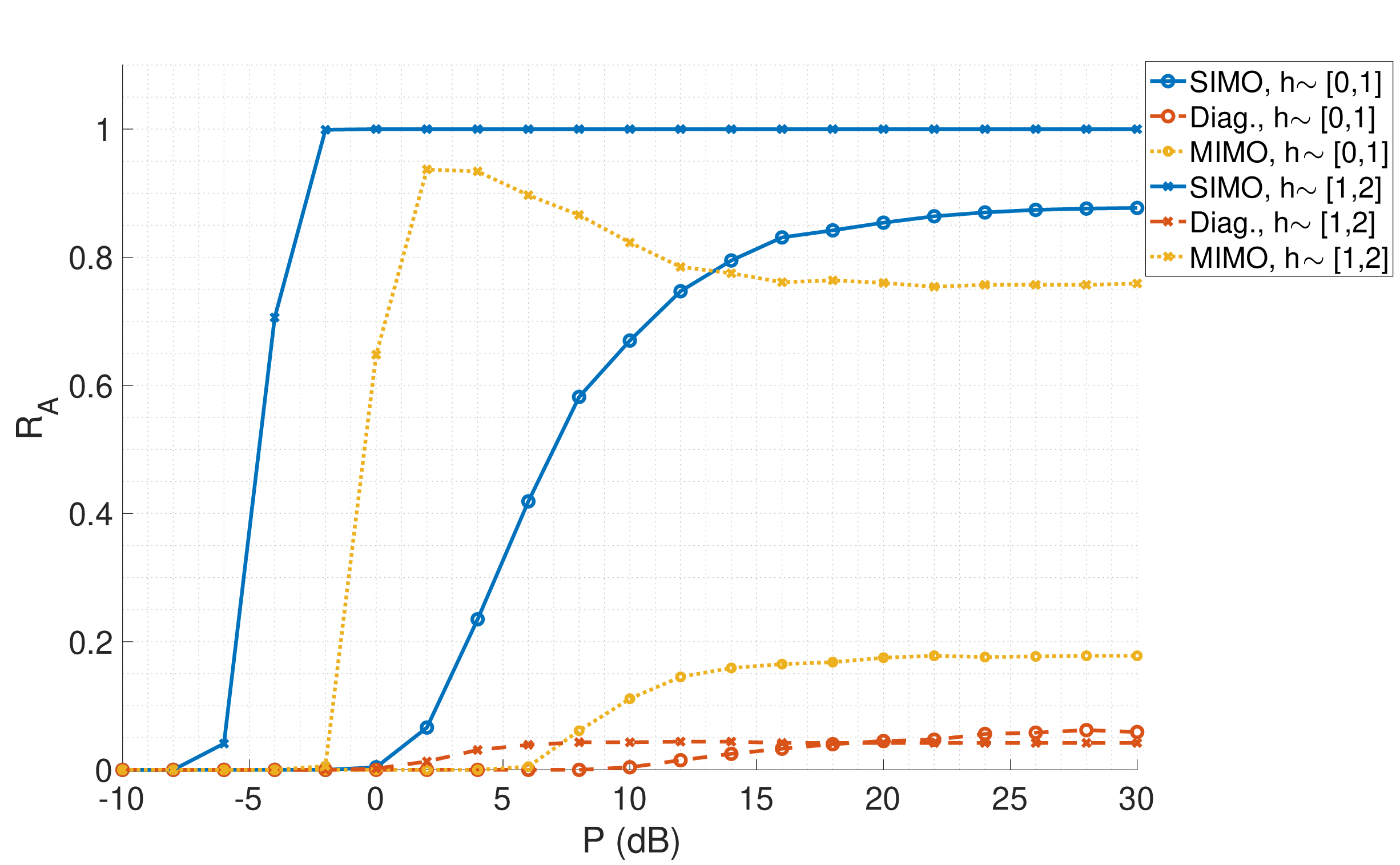}
    \caption{Simulation results for SIMO MAC and $2$-by-$2$ MIMO MAC with CFMA-SCS where the channel coefficients are generated from Uniform $[0,1]$ and Uniform $[1,2]$. It is harder for CFMA-SCS to achieve the sum capacity in the MIMO case compared to the SIMO case. When the channel matrices are diagonal in the MIMO case, CFMA-SCS has a very poor sum capacity achievability. In the SIMO MAC case where the channel coefficients are generated from Uniform $[1,2]$, CFMA-SCS can achieve the sum capacity when the SNR is larger than $0$~dB.}
    \label{fig_SIMO_MIMO}
\end{figure}

We plot $R_A$ against the power constraint $P$ in Fig.~\ref{fig_SIMO_MIMO} for the discussed channel models. The solid curve refers to the SIMO case with $r=2$. The dashed and dotted curves refer to the $2$-by-$2$ MIMO case with diagonal and generic channels, respectively. We use two distributions to generate the channel coefficients, i.e., Uniform $[0,1]$ and Uniform $[1,2]$. They are distinguished by the circle and cross mark in Fig.~\ref{fig_SIMO_MIMO}. We observe that in general it is easier for CFMA-SCS to achieve the sum capacity in the SIMO cases compared to the MIMO cases. Especially, $R_A$ achieves $1$ when $P>0$~dB in the SIMO case when the channel coefficients are generated from Uniform $[1,2]$. On the other hand, $R_A$ is extremely small (maximum around $0.05$) in the diagonal MIMO cases. This result is reasonable because we have shown in Sec.~\ref{sec_2by2_diagonal}) that the sum capacity achieving conditions for diagonal channels are very strict. In the generic MIMO case, the sum capacity achievability highly depends on the distribution where the channel coefficients generated from. We can observe that $R_A$ is always below $0.2$ when the channel coefficients are generated from Uniform $[0,1]$. However, $R_A$ can keep over $0.75$ for $P>1$~dB and achieves over $0.93$ when $2$~dB $<P<4$~dB. This also implies that a large power may not benefit CFMA-SCS to achieve the sum capacity. 

\subsection{Permutation encoding in CFMA-SCS}\label{sec_eg_perm}
In this subsection, we will investigate the case where the precoding matrices $\textbf{B}_l$ can be chosen differently from the Cholesky decomposition of the input covariance matrices $\textbf{K}_l$ for $l=1,2$. Recall that in the encoding process, we apply the precoding matrices such that $\textbf{B}_l\textbf{B}_l^T = \textbf{K}_l$. Although $\textbf{B}_l$ resulting from the Cholesky decomposition is a valid choice, there are different decomposition of $\textbf{K}_l$, which may result in different rate pair, hence affecting the capacity achievability. To show this, we introduce the permutation matrices $\textbf{P}_1,\textbf{P}_2$. A permutation matrix is a square binary matrix that has exactly one entry of $1$ in each row
and each column with all other entries being $0$. From Theorem~\ref{thm_sum_capacity}, the precoding matrices $\textbf{B}_l^*$ should satisfy $\textbf{B}_l^*\textbf{B}_l^{*T} = \textbf{K}_l^*$ for $l=1,2$ to achieve the sum capacity. We call the construction as SCS without permutation when $\textbf{B}_l^*$ is derived from the Cholesky decomposition of $\textbf{K}_l^*$. We also construct the new precoding matrices $\tilde{\textbf{B}}_l^* = \textbf{B}_l^*\textbf{P}_l$ for $l=1,2$, where this construction is called SCS with permutation. Note that both constructions satisfy the input covariance matrices since $\textbf{P}_l$ are orthogonal, i.e., $\textbf{P}_l\textbf{P}_l^T = \textbf{I}$.

\begin{figure}
    \centering
    \includegraphics[width = 0.85\linewidth]{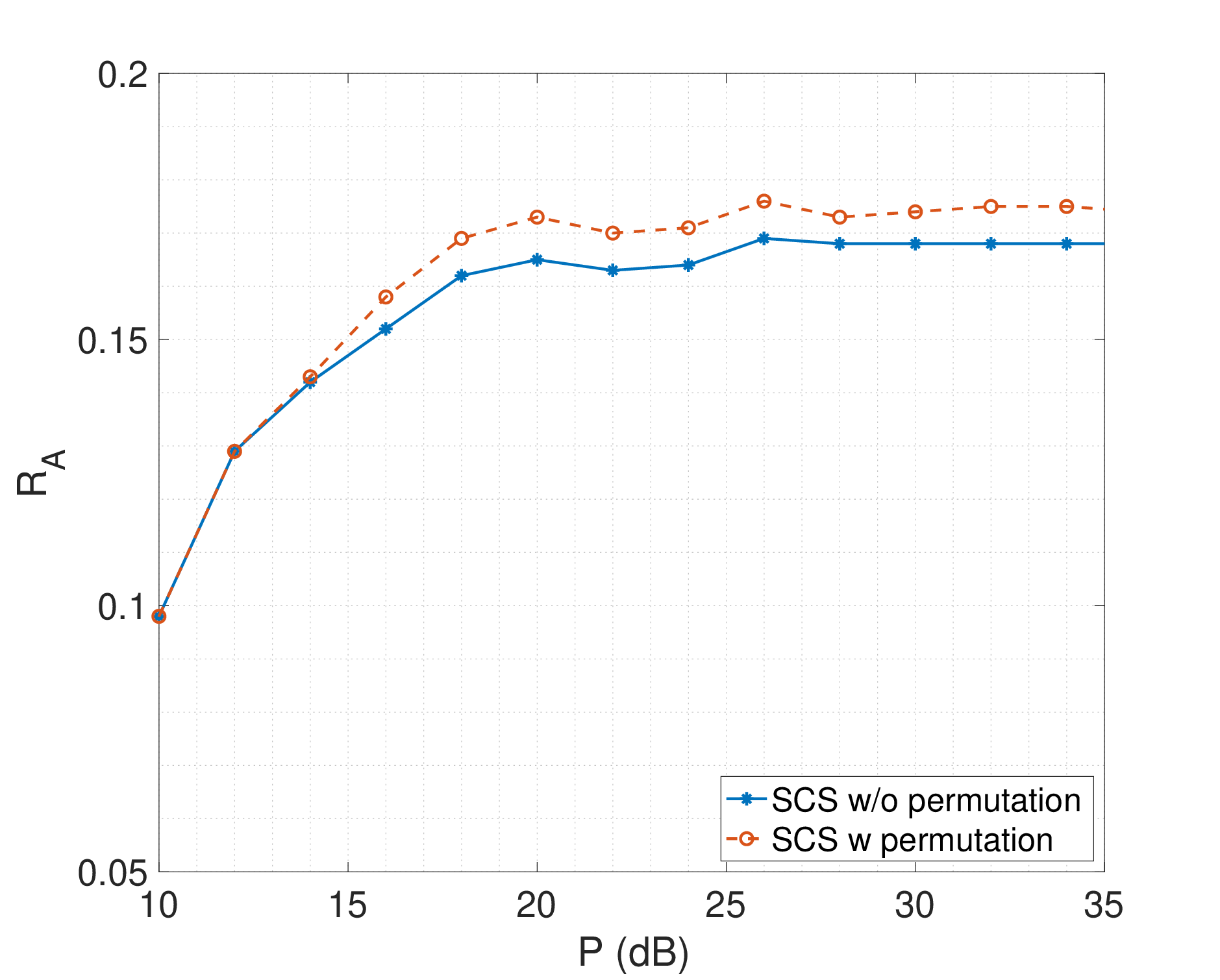}
    \caption{Simulation results for a two-user generic $2$-by-$2$ MIMO MAC with CFMA-SCS, where the channel coefficients are randomly distributed over the interval $[0,1]$. There is a nearly constant improvement (around $0.01$) of $R_A$ with permutation when the power constraint $P$ is larger than $20$~dB.}
    \label{fig_MIMO_Uniform01_Perm}
\end{figure}

We conduct the numerical studies with generic $2$-by-$2$ MAC channels, where each user and the receiver have two antennas. Each channel coefficient in the $2$-by-$2$ channel matrices $\textbf{H}_1,\textbf{H}_2$ is randomly generated from a uniform distribution over the interval $[0,1]$, which is the same as Sec.~\ref{sec_eg}. We generate multiple channel realizations and check the sum capacity achievability for each realization. In this case, we will study $R_A$ for different power constraints. Recall that $R_A$ is the ratio of the amount of the channel realizations whose sum capacity can be achieved on the total amount of the generated channel realizations. For CFMA-SCS without permutation, the simulation is the same as Sec.~\ref{sec_eg}, while for CFMA-SCS with permutation we replace the precoding matrices with $\textbf{B}_l\textbf{P}_l$ for $l=1,2$ and check the sum capacity achievability with all possible permutation matrices $\textbf{P}_1,\textbf{P}_2$. The results are shown in Fig.~\ref{fig_MIMO_Uniform01_Perm}. When the power constraint $P$ is below $12$~dB, $R_A$ has no obvious change when permutation is introduced. However, when $P$ is above $15$~dB, CFMA-SCS with permutation has a larger $R_A$. The improvement of $R_A$ is around $0.01$ when $P$ is larger than $20$~dB.

\subsection{Comparison between CFMA-SCS and CFMA-PCS}\label{sec_eg_individual_codebook}
We first look at a two-user $2$-by-$2$ MIMO MAC with channel matrices 
$$\textbf{H}_1 = \left[\begin{array}{cc}
    1.3 & 1.2 \\
    1.3 & 1.8
\end{array}\right],\;
\textbf{H}_2 = \left[\begin{array}{cc}
    1.4 & 1.2 \\
    1.2 & 1.9
\end{array}\right].$$

We check whether the sum capacity can be achieved with both CFMA-SCS and CFMA-PCS when the power constraint $P$ varies from $0$~dB to $24$~dB. CFMA-SCS is tested using Theorem~\ref{thm_sum_capacity} while CFMA-PCS is tested with Theorem~\ref{thm_capacity_achieve_individual_codebook}. The results are given in Table~\ref{tab_comparison}, where ``$\times$'' means the sum capacity can not be achieved while ``$\checkmark$'' means it is achievable.  
\begin{table}
    \centering
    \caption{Sum capacity achievability for a given MIMO MAC.}
    \label{tab_comparison}
    \begin{tabular}{|c|c|c|c|c|c|c|c|c|c|c|c|c|c|}
        \hline
        P (dB) & 0 & 2 & 4 & 6 & 8 & 10 & 12 & 14 & 16 & 18 & 20 & 22 & 24\\
        \hline
        CFMA-SCS & $\checkmark$ & $\checkmark$ & $\checkmark$ & $\times$ & $\times$ & $\times$ & $\times$ & $\times$ & $\times$ & $\times$ & $\times$ & $\times$ & $\times$ \\
        \hline
        CFMA-PCS & $\checkmark$ & $\checkmark$ & $\checkmark$ & $\checkmark$ & $\times$ & $\checkmark$ & $\checkmark$ & $\checkmark$ & $\checkmark$ & $\checkmark$ & $\checkmark$ & $\times$ & $\times$\\
        \hline
    \end{tabular}    
\end{table}

To investigate what type of channel and power can benefit the sum capacity achievability, we look at the singular value decomposition of the matrices $\textbf{H}_l\textbf{B}_l^*$ for $l=1,2$. Note that $\textbf{B}_l^*$ is defined in Sec.~\ref{sec_sum_capacity_achievability_SCS} and depends on both the channel matrices $\textbf{H}_l$ and the power constraint $P$. When $P=0$~dB, we have $\textbf{B}_1^* = \left[\begin{array}{cc}
    0.636 & 0 \\
    0.772 & 0
\end{array}\right]$, $\textbf{B}_2^* = \left[\begin{array}{cc}
    0.646 & 0 \\
    0.763 & 0
\end{array}\right]$. With SVD, we can find $\textbf{H}_l\textbf{B}_l^* = \textbf{S}_l\textbf{V}_l\textbf{D}_l$ for $l=1,2$ where 
\begin{align*}
    & \textbf{S}_1 = \left[\begin{array}{cc}
    -0.62 & -0.784 \\
   -0.784  &  0.62
\end{array}\right],\textbf{V}_1 = \left[\begin{array}{cc}
    2.825 & 0 \\
    0 & 0
\end{array}\right],\textbf{D}_1 = \left[\begin{array}{cc}
    -1 &  0 \\
    0  &  1
\end{array}\right] \\
& \textbf{S}_2 = \left[\begin{array}{cc}
    -0.633 &  -0.774 \\
    -0.774  &  0.633
\end{array}\right],\textbf{V}_2 = \left[\begin{array}{cc}
    2.875 & 0 \\
    0 & 0
\end{array}\right],\textbf{D}_2 = \left[\begin{array}{cc}
    -1 &  0 \\
    0  &  1
\end{array}\right]
\end{align*}
It can be observed that $\textbf{S}_1$ and $\textbf{S}_2$ are very similar while $\textbf{D}_1$ and $\textbf{D}_2$ are the same. This is approximately the special case we discussed in Sec.~\ref{sec_decomposition}. In this case, the sum capacity is achievable with CFMA-SCS. We can also observe that the first singular value of each channel is $\lambda_{11} = 1.75$ and $\lambda_{21} = 1.82$, respectively. They are both larger than $\sqrt{3/2}$, which is consistent with Remark~\ref{remark_decomposition}. Similar results hold when $P = 2$~dB and $4$~dB, where with CFMA-SCS the sum capacity is achievable. When $P$ has other values, the SVD is not as the form as the previous cases and the sum capacity is not achievable with CFMA-SCS. It seems that the channel matrices of the special case in Sec.~\ref{sec_decomposition} will benefit the sum capacity achievability of CFMA-SCS. However, the sum capacity achievability of CFMA-PCS does not depend directly on this special channel form. However, CFMA-PCS can achieve the sum capacity for more power constraints than CFMA-SCS in this channel condition. Another interesting observation is that a large power is not beneficial for the sum capacity achievability of both schemes. For example, when $P > 20$~dB, both schemes cannot achieve the sum capacity.

\begin{figure}
    \centering
    \includegraphics[width = 0.85\linewidth]{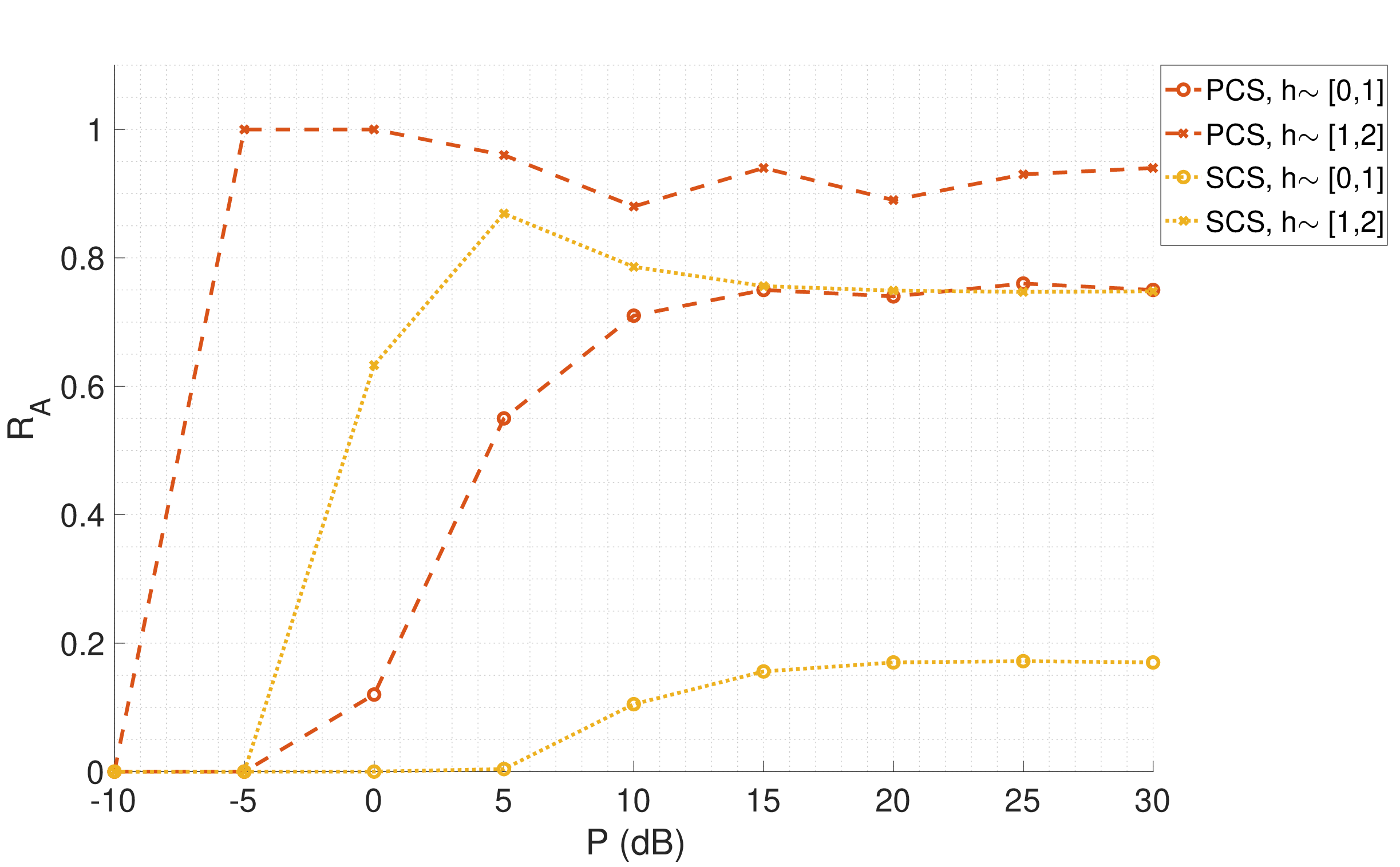}
    \caption{Simulation results for a two-user generic $2$-by-$2$ MIMO MAC, where the channel coefficients are randomly distributed over the interval $[0,1]$ and $[1,2]$. It can be observed that CFMA-PCS has larger $R_A$'s in all cases. However, CFMA-PCS has a larger increase in $R_A$ when the channel coefficients are generated from Uniform $[1,2]$ compared to when they are from Uniform $[0,1]$.}
    \label{fig_MIMO_Uniform01_ind}
\end{figure}

To further compare the sum capacity achievability of both schemes, we perform the same simulations as the previous numerical examples. We generate multiple channel realizations according to given distributions and check whether each channel realization is sum capacity achievable with both schemes. We then plot the ratio $R_A$ against different power constraints. In Fig.~\ref{fig_MIMO_Uniform01_ind}, we generate the channel coefficients from the uniform distribution over the interval $[0,1]$ and $[1,2]$, respectively, which are the same as Sec.~\ref{sec_eg}. The results show that with CFMA-PCS, the sum capacity can be achieved for more channel realizations than that with CFMA-SCS in all cases. This may due to the different ways for codebook constructions. In particular, CFMA-PCS discards the inactive channels due to the independent codebook for each antennas while CFMA-SCS cannot do that. When the channel coefficients are generated from Uniform $[1,2]$, both schemes have a larger $R_A$ compared to the case where the channel coefficients are generated from Uniform $[0,1]$. CFMA-SCS seems to be more sensitive to the strength of the line-of-sight (LoS) component.

To summarize the findings from the numerical examples, we have the following comparison for CFMA-SCS and CFMA-PCS.
\begin{itemize}
    \item In general, CFMA-SCS has lower coding complexity than CFMA-PCS because it only requires two codebooks and decodes two linear combinations.
    \item The sum capacity achievability of the two transmission schemes (CFMA-SCS/CFMA-PCS) heavily depend on the channel conditions.
    \item In general, CFMA-PCS has a better sum capacity achievability than CFMA-SCS.
    \item CFMA-SCS has a larger increase in the sum capacity achievability than CFMA-PCS when there is a stronger dominant channel component. 
\end{itemize}

\section{Conclusion}
In this paper, we consider CFMA as the multiple-access technique for the two-user MIMO MAC with both serial and parallel coding schemes. We characterize the achievable rate pair with CFMA-SCS for the general MIMO MAC with an arbitrary number of transmit and receive antennas. We show that it is possible to achieve the sum capacity given proper power and channel conditions. We propose CFMA-PCS and derive the sum capacity achieving conditions using existing SIMO MAC results for compute-forward. Both schemes have advantages and disadvantages. CFMA-SCS requires fewer codebooks and fewer linear combinations to decode while CFMA-PCS has shorter codeword length. The comparison of both schemes is studied for several channel conditions. From the numerical results, CFMA-PCS is better regarding the sum capacity achievability. However, CFMA-SCS still has rooms to improve given that the precoding matrices can be further optimised. Although we consider the two-user case in the paper, it is easy to extend the coding schemes to multiple-user cases by introducing more linear combinations.

\bibliographystyle{IEEEtran}
\bibliography{reference,IEEEabrv}

\begin{thebibliography}{10}
\providecommand{\url}[1]{#1}
\csname url@samestyle\endcsname
\providecommand{\newblock}{\relax}
\providecommand{\bibinfo}[2]{#2}
\providecommand{\BIBentrySTDinterwordspacing}{\spaceskip=0pt\relax}
\providecommand{\BIBentryALTinterwordstretchfactor}{4}
\providecommand{\BIBentryALTinterwordspacing}{\spaceskip=\fontdimen2\font plus
\BIBentryALTinterwordstretchfactor\fontdimen3\font minus \fontdimen4\font\relax}
\providecommand{\BIBforeignlanguage}[2]{{%
\expandafter\ifx\csname l@#1\endcsname\relax
\typeout{** WARNING: IEEEtran.bst: No hyphenation pattern has been}%
\typeout{** loaded for the language `#1'. Using the pattern for}%
\typeout{** the default language instead.}%
\else
\language=\csname l@#1\endcsname
\fi
#2}}
\providecommand{\BIBdecl}{\relax}
\BIBdecl

\bibitem{Tse05fundamentalsof}
D.~Tse and P.~Viswanath, \emph{Fundamentals of Wireless Communication}.\hskip 1em plus 0.5em minus 0.4em\relax Cambridge University Press, 2005.

\bibitem{Zhang06PLNC}
\BIBentryALTinterwordspacing
S.~Zhang, S.~C. Liew, and P.~P. Lam, ``Hot {T}opic: {P}hysical-{L}ayer {N}etwork {C}oding,'' in \emph{Proceedings of the 12th Annual International Conference on Mobile Computing and Networking}, ser. MobiCom '06.\hskip 1em plus 0.5em minus 0.4em\relax New York, NY, USA: Association for Computing Machinery, 2006, p. 358–365. [Online]. Available: \url{https://doi.org/10.1145/1161089.1161129}
\BIBentrySTDinterwordspacing

\bibitem{Nazer11CF}
B.~Nazer and M.~Gastpar, ``Compute-and-{F}orward: {H}arnessing {I}nterference {T}hrough {S}tructured {C}odes,'' \emph{IEEE Transactions on Information Theory}, vol.~57, no.~10, pp. 6463--6486, 2011.

\bibitem{Zhu17}
J.~Zhu and M.~Gastpar, ``Gaussian {M}ultiple {A}ccess via {C}ompute-and-{F}orward,'' \emph{IEEE Transactions on Information Theory}, vol.~63, no.~5, pp. 2678--2695, 2017.

\bibitem{nazer16expand}
B.~Nazer, V.~R. Cadambe, V.~Ntranos, and G.~Caire, ``Expanding the {C}ompute-and-{F}orward {F}ramework: {U}nequal {P}owers, {S}ignal {L}evels, and {M}ultiple {L}inear {C}ombinations,'' \emph{IEEE Transactions on Information Theory}, vol.~62, no.~9, pp. 4879--4909, 2016.

\bibitem{Hong13}
S.-N. Hong and G.~Caire, ``Compute-and-{F}orward {S}trategies for {C}ooperative {D}istributed {A}ntenna {S}ystems,'' \emph{IEEE Transactions on Information Theory}, vol.~59, no.~9, pp. 5227--5243, 2013.

\bibitem{Ordentlich17}
O.~Ordentlich and Y.~Polyanskiy, ``Low {C}omplexity {S}chemes for the {R}andom {A}ccess {G}aussian {C}hannel,'' in \emph{2017 IEEE International Symposium on Information Theory (ISIT)}, 2017, pp. 2528--2532.

\bibitem{Zhan14}
J.~Zhan, B.~Nazer, U.~Erez, and M.~Gastpar, ``Integer-{F}orcing {L}inear {R}eceivers,'' \emph{IEEE Transactions on Information Theory}, vol.~60, no.~12, pp. 7661--7685, 2014.

\bibitem{He18}
W.~He, B.~Nazer, and S.~Shamai~Shitz, ``Uplink-{D}ownlink {D}uality for {I}nteger-{F}orcing,'' \emph{IEEE Transactions on Information Theory}, vol.~64, no.~3, pp. 1992--2011, 2018.

\bibitem{Lyu19}
S.~Lyu, A.~Campello, and C.~Ling, ``Ring {C}ompute-and-{F}orward {O}ver {B}lock-{F}ading {C}hannels,'' \emph{IEEE Transactions on Information Theory}, vol.~65, no.~11, pp. 6931--6949, 2019.

\bibitem{Erez05}
U.~Erez, S.~Litsyn, and R.~Zamir, ``Lattices {W}hich {A}re {G}ood for ({A}lmost) {E}verything,'' \emph{IEEE Transactions on Information Theory}, vol.~51, no.~10, pp. 3401--3416, 2005.

\bibitem{Gamal04}
H.~El~Gamal, G.~Caire, and M.~Damen, ``Lattice {C}oding and {D}ecoding {A}chieve the {O}ptimal {D}iversity-{M}ultiplexing {T}radeoff of {MIMO} {C}hannels,'' \emph{IEEE Transactions on Information Theory}, vol.~50, no.~6, pp. 968--985, 2004.

\bibitem{Erez04}
U.~Erez and R.~Zamir, ``Achieving 1/2 log (1+{SNR}) on the {AWGN} {C}hannel with {L}attice {E}ncoding and {D}ecoding,'' \emph{IEEE Transactions on Information Theory}, vol.~50, no.~10, pp. 2293--2314, 2004.

\bibitem{Hindy15}
A.~Hindy and A.~Nosratinia, ``Achieving the {E}rgodic {C}apacity with {L}attice {C}odes,'' in \emph{2015 IEEE International Symposium on Information Theory (ISIT)}, 2015, pp. 441--445.

\bibitem{Campello16}
A.~Campello, C.~Ling, and J.-C. Belfiore, ``Algebraic {L}attices {A}chieving the {C}apacity of the {E}rgodic {F}ading {C}hannel,'' in \emph{2016 IEEE Information Theory Workshop (ITW)}, 2016, pp. 459--463.

\bibitem{Ordentlich15}
O.~Ordentlich and U.~Erez, ``Precoded {I}nteger-{F}orcing {U}niversally {A}chieves the {MIMO} {C}apacity to {W}ithin a {C}onstant {G}ap,'' \emph{IEEE Transactions on Information Theory}, vol.~61, no.~1, pp. 323--340, 2015.

\bibitem{Lin11}
S.-C. Lin, P.-H. Lin, C.-P. Lee, and H.-J. Su, ``Filter and {N}ested {L}attice {C}ode {D}esign for {MIMO} {F}ading {C}hannels with {S}ide-{I}nformation,'' \emph{IEEE Transactions on Communications}, vol.~59, no.~6, pp. 1489--1494, 2011.

\bibitem{Loeliger97}
H.-A. Loeliger, ``Averaging {B}ounds for {L}attices and {L}inear {C}odes,'' \emph{IEEE Transactions on Information Theory}, vol.~43, no.~6, pp. 1767--1773, 1997.

\bibitem{gamal_kim_2011}
A.~El~Gamal and Y.-H. Kim, \emph{Network Information Theory}.\hskip 1em plus 0.5em minus 0.4em\relax Cambridge University Press, 2011.

\bibitem{Basu16AlgorithmAlgebra}
S.~Basu, R.~Pollack, and M.-F. Roy, \emph{Algorithms in Real Algebraic Geometry}.\hskip 1em plus 0.5em minus 0.4em\relax Springer, 2016.

\bibitem{Andrilli23LinearAlgebra}
S.~Andrilli and D.~Hecker, \emph{Elementary {L}inear {A}lgebra}.\hskip 1em plus 0.5em minus 0.4em\relax Academic Press, 2023.

\bibitem{Sylvester1851}
J.~J. Sylvester, ``On the {R}elation {B}etween the {M}inor {D}eterminants of {L}inearly {E}quivalent {Q}uadratic {F}unctions,'' \emph{Philosophical Magazine}, vol.~1, pp. 295--305, 1851.

\end{thebibliography}

\end{document}